\pdfoutput=1
\documentclass[12pt,a4paper]{article}

\usepackage{ifthen} 
\newboolean{pdflatex}
\setboolean{pdflatex}{true} 

\newboolean{articletitles}
\setboolean{articletitles}{true} 

\newboolean{uprightparticles}
\setboolean{uprightparticles}{false} 

\def\paperauthors{LHCb collaboration} 
\def\paperasciititle{A study of CP violation in the decays B2DhD2KKpipi and B2DhD2pipipipi} 
\def\papertitle{A study of \CP violation in the decays $\Bpm\to[\Kp\Km\pip\pim]_\D h^{\pm}$ ($h = K, \pi$) and $\Bpm\to[\pip\pim\pip\pim]_\D h^{\pm}$} 
\def\paperkeywords{{High Energy Physics}, {LHCb}} 
\def\papercopyright{\the\year\ CERN for the benefit of the LHCb collaboration} 
\def\paperlicence{CC BY 4.0 licence}
\def\paperlicenceurl{https://creativecommons.org/licenses/by/4.0/}

\newcommand*\diff{\mathop{}\!\mathrm{d}}

\usepackage{booktabs}

\usepackage{subcaption}

\usepackage[top=1in, bottom=1.25in, left=1in, right=1in]{geometry}

\usepackage{microtype}
\usepackage{lineno}  
\usepackage{xspace} 
\usepackage{caption} 

\usepackage{graphicx}  
\usepackage{color}
\usepackage{colortbl}
\graphicspath{{./figs/}} 

\usepackage{amsmath} 
\usepackage{amssymb}
\usepackage{amsfonts}
\usepackage{upgreek} 

\newcommand*\patchAmsMathEnvironmentForLineno[1]{%
\expandafter\let\csname old#1\expandafter\endcsname\csname #1\endcsname
\expandafter\let\csname oldend#1\expandafter\endcsname\csname
end#1\endcsname
 \renewenvironment{#1}%
   {\linenomath\csname old#1\endcsname}%
   {\csname oldend#1\endcsname\endlinenomath}%
}
\newcommand*\patchBothAmsMathEnvironmentsForLineno[1]{%
  \patchAmsMathEnvironmentForLineno{#1}%
  \patchAmsMathEnvironmentForLineno{#1*}%
}
\AtBeginDocument{%
\patchBothAmsMathEnvironmentsForLineno{equation}%
\patchBothAmsMathEnvironmentsForLineno{align}%
\patchBothAmsMathEnvironmentsForLineno{flalign}%
\patchBothAmsMathEnvironmentsForLineno{alignat}%
\patchBothAmsMathEnvironmentsForLineno{gather}%
\patchBothAmsMathEnvironmentsForLineno{multline}%
\patchBothAmsMathEnvironmentsForLineno{eqnarray}%
}

\usepackage{hyperxmp}

\usepackage[pdftex,
            pdfauthor={\paperauthors},
            pdftitle={\paperasciititle},
            pdfkeywords={\paperkeywords},
            pdfcopyright={Copyright (C) \papercopyright},
            pdflicenseurl={\paperlicenceurl}]{hyperref}

\usepackage[all]{hypcap} 

\usepackage{xspace} 
\usepackage{upgreek}

\def\lhcb   {\mbox{LHCb}\xspace}

\def\cleo   {\mbox{CLEO}\xspace}

\def\MagUp {\mbox{\em Mag\kern -0.05em Up}\xspace}

\ifthenelse{\boolean{uprightparticles}}%
{

 \def\Ppi         {\ensuremath{\uppi}\xspace}

 \def\PDelta      {\ensuremath{\Delta}\xspace}                 
 \def\PXi         {\ensuremath{\Xi}\xspace}                 
 \def\PLambda     {\ensuremath{\Lambda}\xspace}                 
 \def\PSigma      {\ensuremath{\Sigma}\xspace}                 
 \def\POmega      {\ensuremath{\Omega}\xspace}                 
 \def\PUpsilon    {\ensuremath{\Upsilon}\xspace}
 \let\oldPi\Pi
 \def\PPi         {\ensuremath{\oldPi}\xspace}

 \def\PB      {\ensuremath{\mathrm{B}}\xspace}                 
                  
 \def\PD      {\ensuremath{\mathrm{D}}\xspace}

 \def\PK      {\ensuremath{\mathrm{K}}\xspace}

 \def\Pb      {\ensuremath{\mathrm{b}}\xspace}                 
 \def\Pc      {\ensuremath{\mathrm{c}}\xspace}                 
 \def\Pd      {\ensuremath{\mathrm{d}}\xspace}

 \def\Pi      {\ensuremath{\mathrm{i}}\xspace}

 \def\Pp      {\ensuremath{\mathrm{p}}\xspace}

 \def\Ps      {\ensuremath{\mathrm{s}}\xspace}                 
                  
 \def\Pu      {\ensuremath{\mathrm{u}}\xspace}

 \def\thebaroffset{0.0em}
}
{

 \def\Ppi         {\ensuremath{\pi}\xspace}

 \mathchardef\PDelta="7101
 \mathchardef\PXi="7104
 \mathchardef\PLambda="7103
 \mathchardef\PSigma="7106
 \mathchardef\POmega="710A
 \mathchardef\PUpsilon="7107
 \mathchardef\PPi="7105
                  
 \def\PB      {\ensuremath{B}\xspace}                 
                  
 \def\PD      {\ensuremath{D}\xspace}

 \def\PK      {\ensuremath{K}\xspace}

 \def\Pb      {\ensuremath{b}\xspace}                 
 \def\Pc      {\ensuremath{c}\xspace}                 
 \def\Pd      {\ensuremath{d}\xspace}

 \def\Pi      {\ensuremath{i}\xspace}

 \def\Pp      {\ensuremath{p}\xspace}

 \def\Ps      {\ensuremath{s}\xspace}                 
                  
 \def\Pu      {\ensuremath{u}\xspace}

 \def\thebaroffset{0.18em}
}
\newcommand{\offsetoverline}[2][\thebaroffset]{\kern #1\overline{\kern -#1 #2}}%

\makeatletter
\ifcase \@ptsize \relax
  \newcommand{\miniscule}{\@setfontsize\miniscule{4}{5}}
\or
  \newcommand{\miniscule}{\@setfontsize\miniscule{5}{6}}
\or
  \newcommand{\miniscule}{\@setfontsize\miniscule{5}{6}}
\fi
\makeatother

\DeclareRobustCommand{\optbar}[1]{\shortstack{{\miniscule (\rule[.5ex]{1.25em}{.18mm})}
  \\ [-.7ex] $#1$}}

\def\uquark    {{\ensuremath{\Pu}}\xspace}

\def\dquark    {{\ensuremath{\Pd}}\xspace}

\def\squark    {{\ensuremath{\Ps}}\xspace}

\def\cquark    {{\ensuremath{\Pc}}\xspace}

\def\bquark    {{\ensuremath{\Pb}}\xspace}

\def\pion   {{\ensuremath{\Ppi}}\xspace}
\def\piz    {{\ensuremath{\pion^0}}\xspace}
\def\pip    {{\ensuremath{\pion^+}}\xspace}
\def\pim    {{\ensuremath{\pion^-}}\xspace}
\def\pipm   {{\ensuremath{\pion^\pm}}\xspace}
\def\pimp   {{\ensuremath{\pion^\mp}}\xspace}

\def\kaon    {{\ensuremath{\PK}}\xspace}

\def\KorKbar {\kern \thebaroffset\optbar{\kern -\thebaroffset \PK}{}\xspace}

\def\Kp      {{\ensuremath{\kaon^+}}\xspace}
\def\Km      {{\ensuremath{\kaon^-}}\xspace}
\def\Kpm     {{\ensuremath{\kaon^\pm}}\xspace}
\def\Kmp     {{\ensuremath{\kaon^\mp}}\xspace}
\def\KS      {{\ensuremath{\kaon^0_{\mathrm{S}}}}\xspace}

\def\Dbar    {{\ensuremath{\offsetoverline{\PD}}}\xspace}
\def\D       {{\ensuremath{\PD}}\xspace}

\def\DorDbar {\kern \thebaroffset\optbar{\kern -\thebaroffset \PD}\xspace}
\def\Dz      {{\ensuremath{\D^0}}\xspace}
\def\Dzb     {{\ensuremath{\Dbar{}^0}}\xspace}
\def\Dp      {{\ensuremath{\D^+}}\xspace}
\def\Dm      {{\ensuremath{\D^-}}\xspace}

\def\DpDm    {\ensuremath{\Dp {\kern -0.16em \Dm}}\xspace}

\def\B       {{\ensuremath{\PB}}\xspace}
\def\Bbar    {{\ensuremath{\offsetoverline{\PB}}}\xspace}

\def\BorBbar {\kern \thebaroffset\optbar{\kern -\thebaroffset \PB}\xspace}

\def\Bd      {{\ensuremath{\B^0}}\xspace}

\def\BdorBdbar {\kern \thebaroffset\optbar{\kern -\thebaroffset \Bd}\xspace}
\def\Bu      {{\ensuremath{\B^+}}\xspace}
\def\Bub     {{\ensuremath{\B^-}}\xspace}
\def\Bp      {{\ensuremath{\Bu}}\xspace}
\def\Bm      {{\ensuremath{\Bub}}\xspace}
\def\Bpm     {{\ensuremath{\B^\pm}}\xspace}

\def\Bs      {{\ensuremath{\B^0_\squark}}\xspace}
\def\Bsb     {{\ensuremath{\Bbar{}^0_\squark}}\xspace}
\def\BsorBsbar {\kern \thebaroffset\optbar{\kern -\thebaroffset \Bs}\xspace}

\def\Y#1S{\ensuremath{\PUpsilon{(#1S)}}\xspace}

\def\proton      {{\ensuremath{\Pp}}\xspace}

\def\Lz          {{\ensuremath{\PLambda}}\xspace}

\def\LorLbar     {\kern \thebaroffset\optbar{\kern -\thebaroffset \PLambda}\xspace}

\def\Lb           {{\ensuremath{\Lz^0_\bquark}}\xspace}

\def\to                 {\ensuremath{\rightarrow}\xspace}

\def\CP                {{\ensuremath{C\!P}}\xspace}

\def\Vud  {{\ensuremath{V_{\uquark\dquark}^{\phantom{\ast}}}}\xspace}
\def\Vcd  {{\ensuremath{V_{\cquark\dquark}^{\phantom{\ast}}}}\xspace}

\def\Vubs  {{\ensuremath{V_{\uquark\bquark}^\ast}}\xspace}
\def\Vcbs  {{\ensuremath{V_{\cquark\bquark}^\ast}}\xspace}

\def\AT#1     {\ensuremath{A_{\mathrm{T}}^{#1}}\xspace}           

\def\C#1      {\ensuremath{\mathcal{C}_{#1}}\xspace}                       
\def\Cp#1     {\ensuremath{\mathcal{C}_{#1}^{'}}\xspace}                    
\def\Ceff#1   {\ensuremath{\mathcal{C}_{#1}^{\mathrm{(eff)}}}\xspace}        
\def\Cpeff#1  {\ensuremath{\mathcal{C}_{#1}^{'\mathrm{(eff)}}}\xspace}       
\def\Ope#1    {\ensuremath{\mathcal{O}_{#1}}\xspace}                       
\def\Opep#1   {\ensuremath{\mathcal{O}_{#1}^{'}}\xspace}                    

\newcommand{\nospaceunit}[1]{\ensuremath{\text{#1}}}       
\newcommand{\aunit}[1]{\ensuremath{\text{\,#1}}}       

\newcommand{\tev}{\aunit{Te\kern -0.1em V}\xspace}
\newcommand{\gev}{\aunit{Ge\kern -0.1em V}\xspace}
\newcommand{\mev}{\aunit{Me\kern -0.1em V}\xspace}
\newcommand{\kev}{\aunit{ke\kern -0.1em V}\xspace}
\newcommand{\ev}{\aunit{e\kern -0.1em V}\xspace}
 
\newcommand{\mevc}{\ensuremath{\aunit{Me\kern -0.1em V\!/}c}\xspace}
\newcommand{\gevc}{\ensuremath{\aunit{Ge\kern -0.1em V\!/}c}\xspace}
\newcommand{\mevcc}{\ensuremath{\aunit{Me\kern -0.1em V\!/}c^2}\xspace}
\newcommand{\gevcc}{\ensuremath{\aunit{Ge\kern -0.1em V\!/}c^2}\xspace}

\def\mum  {\ensuremath{\,\upmu\nospaceunit{m}}\xspace}

\def\fb   {\ensuremath{\aunit{fb}}\xspace}
\def\invfb   {\ensuremath{\fb^{-1}}\xspace}

\def\gsim{{~\raise.15em\hbox{$>$}\kern-.85em
          \lower.35em\hbox{$\sim$}~}\xspace}
\def\lsim{{~\raise.15em\hbox{$<$}\kern-.85em
          \lower.35em\hbox{$\sim$}~}\xspace}

\def\pt         {\ensuremath{p_{\mathrm{T}}}\xspace}

\def\ptot       {\ensuremath{p}\xspace}

\def\evtgen     {\mbox{\textsc{EvtGen}}\xspace}

\def\geant      {\mbox{\textsc{Geant4}}\xspace}

\def\photos     {\mbox{\textsc{Photos}}\xspace}

\def\pythia     {\mbox{\textsc{Pythia}}\xspace}

\def\tell1  {TELL1\xspace}
\def\ukl1   {UKL1\xspace}

\newcommand{\lhcborcid}[1]{\href{https://orcid.org/#1}{\hspace*{0.1em}\raisebox{-0.45ex}{\includegraphics[width=1em]{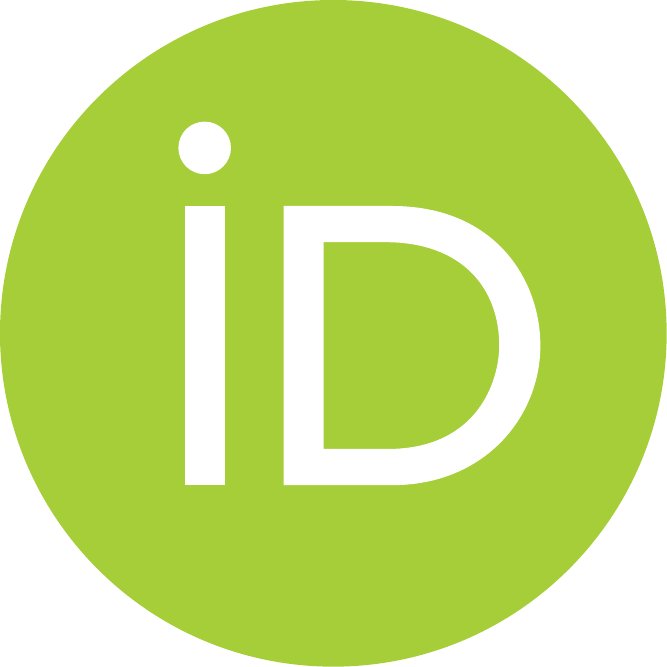}}}}

\usepackage{cite} 
\usepackage{mciteplus}

\begin{document}

\renewcommand{\thefootnote}{\fnsymbol{footnote}}
\setcounter{footnote}{1}


\begin{titlepage}
\pagenumbering{roman}

\vspace*{-1.5cm}
\centerline{\large EUROPEAN ORGANIZATION FOR NUCLEAR RESEARCH (CERN)}
\vspace*{1.5cm}
\noindent
\begin{tabular*}{\linewidth}{lc@{\extracolsep{\fill}}r@{\extracolsep{0pt}}}
\ifthenelse{\boolean{pdflatex}}
{\vspace*{-1.5cm}\mbox{\!\!\!\includegraphics[width=.14\textwidth]{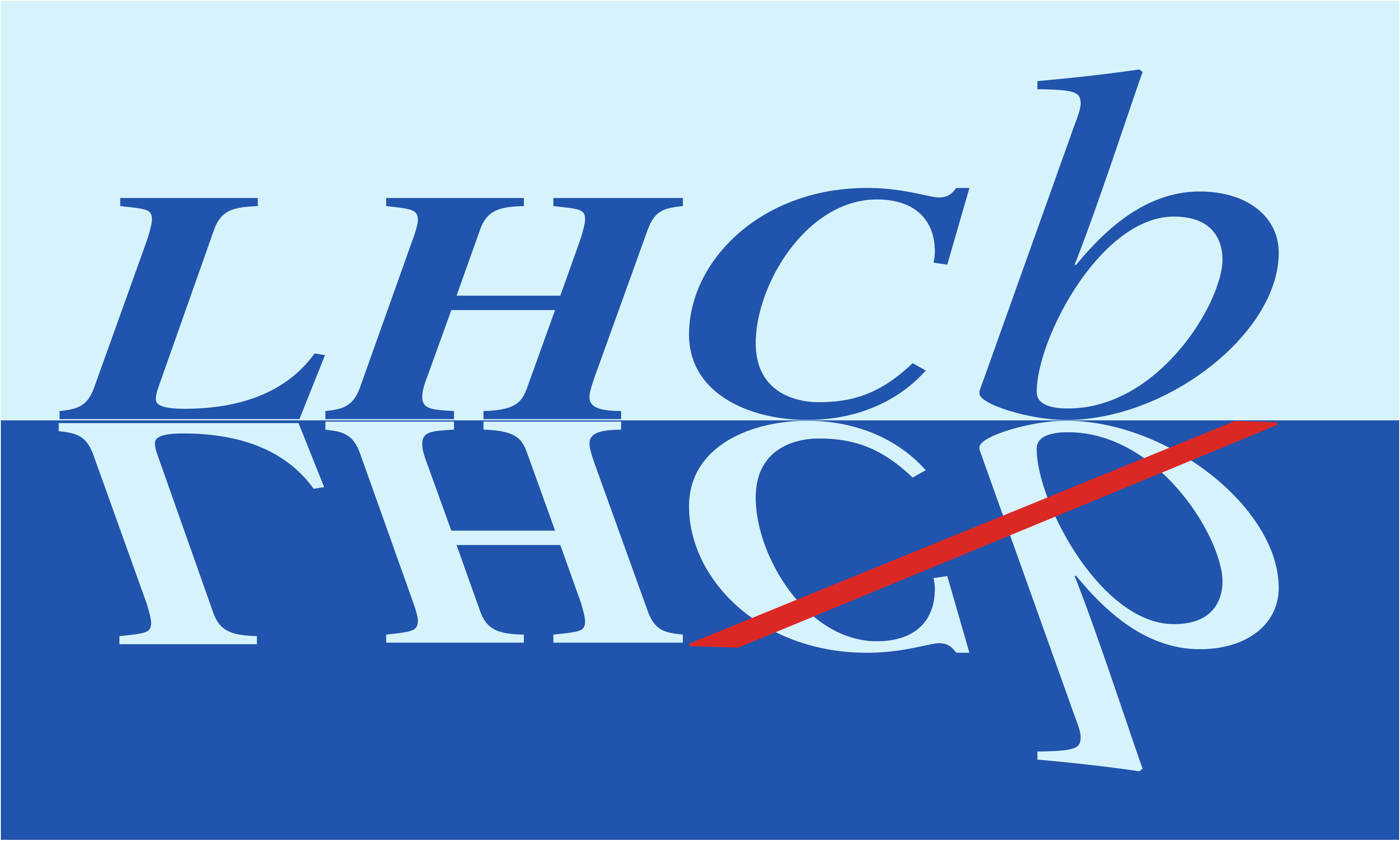}} & &}%
{\vspace*{-1.2cm}\mbox{\!\!\!\includegraphics[width=.12\textwidth]{figs/lhcb-logo.eps}} & &}%
\\
 & & CERN-EP-2022-285 \\  
 & & LHCb-PAPER-2022-037 \\  
 & & June 30, 2023 \\ 
 & & \\
\end{tabular*}

\vspace*{2.0cm}

{\normalfont\bfseries\boldmath\huge
\begin{center}
  \papertitle 
\end{center}
}

\vspace*{2.0cm}

\begin{center}
\paperauthors\footnote{Authors are listed at the end of this paper.}
\end{center}

\vspace{\fill}

\begin{abstract}
  \noindent
  The first study of \CP violation in the decay mode $\Bpm\to[\Kp\Km\pip\pim]_\D h^\pm$, with $h=K,\pi$, is presented, exploiting a data sample of proton-proton collisions collected by the \lhcb experiment that corresponds to an integrated luminosity of $9\invfb$. The analysis is performed in bins of phase space, which are optimised for sensitivity to local \CP asymmetries. \CP-violating observables that are sensitive to the angle $\gamma$ of the Unitarity Triangle are determined. The analysis requires external information on charm-decay parameters, which are currently taken from an amplitude analysis of \lhcb data, but can be updated in the future when direct measurements become available. Measurements are also performed of phase-space integrated observables for $\Bpm\to[\Kp\Km\pip\pim]_\D h^\pm$ and $\Bpm\to[\pip\pim\pip\pim]_\D h^\pm$ decays.
\end{abstract}

\vspace*{2.0cm}

\begin{center}
  Published in
  Eur.~Phys.~J.~C83 547 (2023)
\end{center}

\vspace{\fill}

{\footnotesize 
\centerline{\copyright~\papercopyright. \href{\paperlicenceurl}{\paperlicence}.}}
\vspace*{2mm}

\end{titlepage}


\newpage
\setcounter{page}{2}
\mbox{~}
%
%
%
%


\renewcommand{\thefootnote}{\arabic{footnote}}
\setcounter{footnote}{0}

\cleardoublepage


\pagestyle{plain} 
\setcounter{page}{1}
\pagenumbering{arabic}


\section{Introduction}
\label{section:Introduction}

The Standard Model (SM) description of charge-parity (\CP) violation can be tested by measuring the lengths and angles of the Unitary Triangle, which is a geometrical representation of the Cabibbo–Kobayashi–Maskawa quark-mixing matrix~\cite{Cabibbo:1963yz,Kobayashi:1973fv}. In particular, the \CP-violating phase  $\gamma\equiv\arg(-\Vud\Vubs/\Vcd\Vcbs)$ is the only angle that can be measured at tree level with negligible theoretical uncertainties~\cite{Brod_2014}. Therefore, it makes an excellent SM benchmark that can be compared with other indirect measurements of $\gamma$ that are more likely to be affected by physics beyond the SM.

A powerful decay channel for the measurement of $\gamma$ is $\Bpm\to\D\Kpm$, which proceeds through both favoured $b\to c\bar{u}s$ and a suppressed $b\to u\bar{c}s$ transitions. Interference occurs when the $\D$ meson, which is a superposition of the $\Dz$ and $\Dzb$ states, decays to a final state common to both of these flavour eigenstates. The interference effects are sensitive to $\gamma$, which can in general be determined from measurements of the appropriate \CP  asymmetries and related observables. This strategy has been pursued for a wide range of $\D$ final states at LHCb~\cite{LHCb-PAPER-2016-003,LHCb-PAPER-2019-044,LHCb-PAPER-2020-019,LHCb-PAPER-2020-036,LHCb-PAPER-2021-036} and other $b$-physics experiments~\cite{BaBar:2010uep,cite:KSpipipi0_gamma,Belle:2021efh}. Effects from \CP-violation also occur in the process $\Bpm\to\D\pipm$ through the interference of $b\to c\bar{u}d$ and $b\to u\bar{c}d$ transitions, but these are in general significantly smaller in magnitude.

An interesting class of $\D$ final states are self-conjugate multi-body $\D$ decays. Since the strong-phase difference between the $\Dz$ and $\Dzb\to\Kp\Km\pip\pim$ decays varies across the multi-dimensional phase space of the charm-meson decay, the sensitivity to $\gamma$ is diluted when considering the decay inclusively. However, by performing measurements in suitably chosen localised regions of phase space, the dilution effects can be minimised and the sensitivity to $\gamma$ enhanced~\cite{Bondar, BondarPoluektov2006, BondarPoluektov2008, GiriGrossmanSofferZupan}.

The $\D\to\Kp\Km\pip\pim$ decay mode has been proposed as a promising decay mode for measuring $\gamma$~\cite{cite:RademackerWilkinson}. It has a rich resonance structure and contains only charged particles in the final state, which is advantageous for experiments at a hadron collider. The availability of a detailed amplitude model for this process, based on an analysis of \lhcb data~\cite{LHCb-PAPER-2018-041}, opens up the possibility of identifying those regions of phase space that have high sensitivity to $\gamma$, which then can be probed in an analysis of $\Bpm\to\D\Kpm$ decays. The interpretation of the $\CP$ asymmetries and other observables in these regions requires knowledge of the strong-phase difference between the $\Dz$ and $\Dzb$ meson decays. This knowledge can be obtained from the same amplitude model used to guide the measurements, but it is preferable to take the information from direct determinations made by experiments at charm threshold, such as BESIII~\cite{cite:BESIIIWhitePaper}, as this approach ensures the determination of $\gamma$ has no dependence on model assumptions. An analogous study has recently been performed using the decay mode $D \to\Kpm\pimp\pipm\pimp$~\cite{LHCb-PAPER-2022-017}.

This paper presents the first study of \CP violation in $\Bpm\to[\Kp\Km\pip\pim]_\D h^\pm$ decays, with $h = K, \pi$. The analysis exploits proton-proton ($pp$) collision data collected by LHCb in Runs 1 and 2 of the LHC, corresponding to 9~\invfb of integrated luminosity. The study is performed in localised regions of phase space, defined with guidance from the amplitude model presented in Ref.~\cite{LHCb-PAPER-2018-041}, and the measurements are used to extract a value of $\gamma$, using model predictions for the strong-phase variation in the charm-meson  decay. In addition, first measurements of the global \CP asymmetries are made for this decay and global measurements are updated for the mode $B^\pm \to [\pi^+\pi^-\pi^+\pi^-]_D h^\pm$ with respect to those reported in Ref.~\cite{LHCb-PAPER-2016-003}, which are also interpreted in terms of $\gamma$ and the underlying physics parameters.


\section{Analysis strategy}
\label{section:Analysis_strategy}
This analysis follows the formalism described in Ref.~\cite{LHCb-PAPER-2020-019}. The $\Bm\to[\Kp\Km\pip\pim]_\D\Km$ decay can proceed via the favoured $\Bm\to\Dz\Km$ amplitude, or via the suppressed $\Bm\to\Dzb\Km$ amplitude. The overall amplitude of this decay is a coherent sum of the two decay paths,
\begin{equation}
    \mathcal{A}_\Bm(\Phi) = \mathcal{A}_\Bm^{\Dz\Km}\Big(\mathcal{A}_\Dz(\Phi) + r_\B^{\D\kaon}\exp\big(i(\delta_\B^{\D\kaon} - \gamma)\big)\mathcal{A}_\Dzb(\Phi)\Big),
    \label{equation:Bpm_amplitude}
\end{equation}
where $\mathcal{A}_\Bm^{\Dz\Km}$ is the amplitude of the favoured $\Bm\to\Dz\Km$ decay, $\mathcal{A}_\Dz$ ($\mathcal{A}_\Dzb$) is the amplitude of the $\Dz$ ($\Dzb$) decay, $r_\B^{\D\kaon}$ is the magnitude of the ratio of the $\B$-decay amplitudes and $\delta_\B^{\D\kaon}$ is the strong-phase difference of the amplitudes. Here $\Phi$ labels the position in the five-dimensional phase space of the decay $\Dz\to\Kp\Km\pip\pim$. The corresponding expression for the $\Bp$ decay is obtained by making the substitutions $\gamma\to -\gamma$ and $\mathcal{A}_\Dz\leftrightarrow\mathcal{A}_\Dzb$. The \CP-conjugated amplitude $\mathcal{A}_\Dzb(\Phi)$ is equal to $\mathcal{A}_\Dz(\bar{\Phi})$, where $\bar{\Phi}$ is obtained by swapping the charges and momentum directions of the $\D$-decay products. Here and in subsequent discussion \CP violation in the $\D$-meson system is neglected, which is a good assumption at the current level of experimental sensitivity~\cite{HFLAV18}.

The $\D$-decay phase space is split into $2\times\mathcal{N}$ bins, labelled from $i = -\mathcal{N}$ to $i = \mathcal{N}$, excluding zero. A \CP transformation relates the bins with indices $-i$ and $+i$. The choice of binning scheme is described in Sec.~\ref{section:Binning_scheme}. The expected yield of $\Bm$ decays in bin $i$ is obtained by integrating Eq.~\eqref{equation:Bpm_amplitude}, and the corresponding expression for $\Bp$ decays, over the phase space $\Phi_i$ that belongs to that particular bin. Defining the \CP-violating observables

\begin{equation}
    x_\pm^{\D\kaon} \equiv r_\B^{\D\kaon}\cos(\delta_\B^{\D\kaon}\pm\gamma), \quad y_\pm^{\D\kaon} \equiv r_\B^{\D\kaon}\sin(\delta_\B^{\D\kaon}\pm\gamma),
    \label{equation:CP_observables}
\end{equation}
the yields $N_i^\pm$ of $\Bpm$ candidates in bin $i$ are given by 
\begin{align}
    N_{+i}^+ &= h_\Bp^{\D\kaon}\Big(F_{-i} + \big((x_+^{\D\kaon})^2 + (y_+^{\D\kaon})^2\big)F_{+i} + 2\sqrt{F_{+i}F_{-i}}\big(x_+^{\D\kaon}c_i - y_+^{\D\kaon}s_i\big)\Big), \label{equation:Ni_plus} \\
    N_{-i}^- &= h_\Bm^{\D\kaon}\Big(F_{-i} + \big((x_-^{\D\kaon})^2 + (y_-^{\D\kaon})^2\big)F_{+i} + 2\sqrt{F_{+i}F_{-i}}\big(x_-^{\D\kaon}c_i - y_-^{\D\kaon}s_i\big)\Big), \label{equation:Ni_minus}
\end{align}
where $h_\Bpm^{\D\kaon}$ are normalisation constants. Since $h_\Bp^{\D\kaon}$ and $h_\Bm^{\D\kaon}$ are independent fit parameters, the binned measurement is insensitive to the $\Bpm$ production asymmetry and any charge asymmetry in the detection efficiency of the kaon that accompanies the $\D$ meson.

Equations~\eqref{equation:Ni_plus} and \eqref{equation:Ni_minus} are sensitive to $\gamma$ through the interference terms, and the magnitude of the interference is determined by the size of $r_\B^{\D\kaon}$, which has been measured to be $\approx 0.1$~\cite{LHCb-PAPER-2021-033}. The parameters $F_i$ are defined as
\begin{equation}
    F_i\equiv\frac{\int_i\diff\Phi\eta(\Phi)\lvert\mathcal{A}_\Dz\rvert^2}{\int\diff\Phi\eta(\Phi)\lvert\mathcal{A}_\Dz\rvert^2},
    \label{equation:Fi}
\end{equation}
which are interpreted as the fractional yield of $\Dz$ decays in bin $i$ measured in this analysis. The function $\eta(\Phi)$ accounts for the detection efficiency, which in general depends on the location of the decay in phase space. The strong-phase information is encoded in the parameters $c_i$ and $s_i$, where 
\begin{equation}
    c_i\equiv\frac{\int_i\diff\Phi\lvert\mathcal{A}_\Dz\rvert\lvert\mathcal{A}_\Dzb\rvert\cos(\Delta\delta_\D)}{\sqrt{\int_i\diff\Phi\lvert\mathcal{A}_\Dz\rvert^2\int_i\diff\Phi\lvert\mathcal{A}_\Dzb\rvert^2}},
    \label{equation:ci}
\end{equation}
which is the amplitude-averaged cosine of the strong-phase difference $\Delta\delta_\D = \delta_\D(\Phi) - \delta_\D(\bar{\Phi})$ of the $\D$ decay. The expression for the amplitude-averaged sine of the strong-phase difference $s_i$ is analogous. It follows that $\bar{F}_i = F_{-i}$, $c_{-i} = c_i$ and $s_{-i} = -s_i$, where $\bar{F}_i$ are the corresponding $\Dzb$ fractional bin yields.  The values of $c_i$ and $s_i$ are currently taken from an amplitude model~\cite{LHCb-PAPER-2018-041}. In the future it is expected that direct measurements of $c_i$ and $s_i$ will become available from a sample of correlated $\D\Dbar$ decays collected by the BESIII experiment at charm threshold~\cite{cite:BESIIIWhitePaper}. The anticipated size of this data sample at charm threshold leads to the choice of $\mathcal{N} = 8$ for the number of bins.

Analogous expressions to Eqs.~\ref{equation:CP_observables},~\ref{equation:Ni_plus} and~\ref{equation:Ni_minus} can be written for the decay mode $\Bpm\to\D\pipm$, with the $\D\kaon$ superscripts replaced by $\D\pion$. Since the decay topology is identical to that of $\Bpm\to\D\Kpm$, the phase-space acceptance $\eta(\Phi)$ is expected to be very similar between the two $\Bpm$-decay modes, and studies using simulation samples show that any differences are negligible within the current precision. Thus, the $F_i$ parameters can be considered as common between $\Bpm\to\D\pipm$ and $\Bpm\to\D\Kpm$ decays. The mode $\Bpm\to\D\pipm$ has a branching fraction that is an order of magnitude larger than that of the $\Bpm\to\D\Kpm$ mode, but the interference effects, governed by the parameter $r_\B^{\D\pion}\approx 0.005$~\cite{LHCb-PAPER-2021-033}, are much smaller. Therefore, this decay has a significantly lower sensitivity to $\gamma$, but is a suitable mode for determining the $F_i$ parameters. By including the $\Bpm\to\D\pipm$ channel as a signal mode, the $F_i$ can be treated as free parameters in the analysis, thus the form of the acceptance function $\eta(\Phi)$ is not needed.

From the definition of Eq.~\eqref{equation:Fi}, it follows that $\sum_iF_i = 1$. Therefore, only $2\mathcal{N} - 1$ of the $F_i$ parameters are independent. To accommodate this constraint in the analysis, the $F_i$ parameters are fitted with the alternative parameterisation in terms of the recursive fractions $R_i$,

\begin{equation}
    F_i \equiv
    \begin{cases}
        R_i, & i = -\mathcal{N} \\
        R_i\prod_{j < i}(1 - R_j), & -\mathcal{N} < i < +\mathcal{N} \\
        \prod_{j < i}(1 - R_j), &i = +\mathcal{N}.
    \end{cases}
\end{equation}
Measuring the yields of $\Bpm$ decays in each bin of phase space allows the eight \CP-violating observables $x_\pm^{\D\kaon}$, $y_\pm^{\D\kaon}$, $x_\pm^{\D\pion}$ and $y_\pm^{\D\pion}$ to be determined. These \CP-violating observables can be interpreted in terms of the five underlying physics parameters $\gamma$, $\delta_\B^{\D\kaon}$, $r_\B^{\D\kaon}$, $\delta_\B^{\D\pion}$ and $r_\B^{\D\pion}$. Pseudoexperiments with all eight CP-violating observables as free parameters exhibit unstable fit results for $r_\B^{\D\pion} < 0.03$, due to large correlations between the $\Bpm\to\D\pipm$ \CP-violating observables and the $F_i$ parameters~\cite{LHCb-PAPER-2020-019}. Since $\gamma$ is a common parameter between the $\Bpm\to\D\Kpm$ and $\Bpm\to\D\pipm$ analyses, the \CP-violating observables for the $\Bpm\to\D\pipm$ mode can be re-parameterised as

\begin{equation}
    x_\xi^{\D\pion} = \Re\big(\xi^{\D\pion}\big), \quad y_\xi^{\D\pion} = \Im\big(\xi^{\D\pion}\big), \quad \xi^{\D\pion} = \frac{r_\B^{\D\pion}}{r_\B^{\D\kaon}}\exp\Big(i\big(\delta_\B^{\D\pion} - \delta_\B^{\D\kaon}\big)\Big).
    \label{equation:xi_observables}
\end{equation}

In summary, a measurement of the yields of $\Bpm\to[\Kp\Km\pip\pim]_\D h^\pm$ decays in bins of phase space allows the six \CP-violating observables $x_\pm^{\D\kaon}$, $y_\pm^{\D\kaon}$, $x_\xi^{\D\pion}$ and $y_\xi^{\D\pion}$ to be determined, along with the experiment-specific parameters $F_i$, expressed in terms of $R_i$, and the four normalisation parameters $h_\Bpm^{\D h}$.

Additionally, $\gamma$ can be further constrained by measuring phase-space integrated \CP-violating observables~\cite{Nayak:2014tea,Malde:2015mha}. These observables bring information that is independent of the binned analysis, which is only sensitive to the relative variation in yields across  phase space. In the case of $\D\to\Kp\Km\pip\pim$ decays the phase-space integrated observables are the charge asymmetries
\begin{equation}
    A^{KK\pi\pi}_h \equiv \frac{\Gamma(\Bm\to\D h^-) - \Gamma(\Bp\to\D h^+)}{\Gamma(\Bm\to\D h^-) + \Gamma(\Bp\to\D h^+)},
    \label{equation:Asymmetry}
\end{equation}
for $h = \pi$ or $K$, and the double ratio
\begin{equation}
    R^{KK\pi\pi}_{\CP} \equiv \frac{R_{\kaon\kaon\pion\pion}}{R_{K\pion\pion\pion}}, \quad R_f \equiv \frac{\Gamma(\Bm\to[f]_\D\Km) + \Gamma(\Bp\to[f]_\D\Kp)}{\Gamma(\Bm\to[f]_\D\pim) + \Gamma(\Bp\to[f]_\D\pip)},
    \label{equation:R_CP}
\end{equation}
where in the case of $R_{K\pion\pion\pion}$ the kaon from the $D$-meson decay has the same charge as the pion or kaon from the $B$-meson decay. The value of $R_{K\pion\pion\pion}$ can be determined from the phase-space integrated yields obtained in Ref.~\cite{LHCb-PAPER-2022-017}. By integrating Eqs.~\eqref{equation:Ni_plus} and \eqref{equation:Ni_minus} over all bins $i$, it can be shown that these phase-space integrated \CP-violating observables may be expressed in terms of the underlying physics parameters,
\begin{align}
    A^{KK\pi\pi}_h &= \frac{2r_\B^{\D h}\kappa\sin(\delta_\B^{\D h})\sin(\gamma)}{1 + (r_\B^{\D h})^2 + 2r_\B^{\D h}\kappa\cos(\delta_\B^{\D h})\cos(\gamma)}, \label{equation:Asymmetry_gamma} \\[0.5em]
    R^{KK\pi\pi}_{\CP} &= 1 + (r_\B^{\D\kaon})^2 + 2r_\B^{\D\kaon}\kappa\cos(\delta_\B^{\D\kaon})\cos(\gamma), \label{equation:R_CP_gamma}
\end{align}
where $\kappa = 2F^{KK\pi\pi}_+ - 1$ is the dilution factor when integrating over all of phase space, with $F^{KK\pi\pi}_+$ being the \CP-even fraction of the decay.  In Eqs.~\eqref{equation:Asymmetry_gamma} and \eqref{equation:R_CP_gamma}, the small effects of charm mixing have been neglected, but these may readily be included~\cite{Rama:2013voa}. Analogous observables $A^{\pi\pi\pi\pi}_h$ and $R^{\pi\pi\pi\pi}_{\CP}$ exist for the decay $D \to \pip\pim\pip\pim$.


\section{Binning scheme}
\label{section:Binning_scheme}
In the $\Bpm\to\D\Kpm$, $\D\to\KS h^+h^-$ ($h=\pion,\kaon$) analysis presented in Ref.~\cite{LHCb-PAPER-2020-019}, an optimal binning scheme, which can be visualised in a two-dimensional Dalitz plot, is used to maximise the sensitivity to $\gamma$. The binning scheme for the $\Dz\to\Kp\Km\pip\pim$ decay is defined analogously, but cannot be easily visualised, as the phase space of four-body decays is five-dimensional.

When defining a binning scheme for the decay $\D \to\Kp\Km\pip\pim$, there are two main requirements. Firstly, it should minimise the dilution of the strong phases when averaged over each bin. Secondly, it should maximise the interference effects and thus the sensitivity to $\gamma$. The scheme is constructed with the guidance   of the amplitude model presented in Ref.~\cite{LHCb-PAPER-2018-041}.
 
The position of a $\D$ decay in phase space is specified by the four-momenta of the decay products.  These four-momenta allow the corresponding decay amplitudes $\mathcal{A}_\Dz$ and $\mathcal{A}_\Dzb$ of the $\Dz$ and $\Dzb$ decays, respectively, to be determined from the amplitude model.  From the $\Dz$ and $\Dzb$ decay amplitudes, two convenient parameters,

\begin{equation}
    \Delta\delta_\D\equiv\arg\Big(\frac{\mathcal{A}_\Dz}{\mathcal{A}_{\Dzb}}\Big), \quad r_\D\equiv\Big\lvert\frac{\mathcal{A}_\Dz}{\mathcal{A}_{\Dzb}}\Big\rvert,
\end{equation}
are defined. These are the strong-phase difference and the magnitude of the ratio between the $\Dz$ and $\Dzb$ amplitudes according to the model, respectively. Effectively, by considering $\Delta\delta_\D$ and $r_\D$ as parameters, each $\D$ decay in the five-dimensional phase space is projected onto a two-dimensional surface, where a binning scheme can be defined.

The binning is first performed in the $\Delta\delta_\D$ coordinate, which spans the range $[-\pi, \pi]$ and is divided into $\mathcal{N}$ bins with boundaries that are symmetric around $\Delta\delta_\D=0$. Assigning a decay to a particular bin ensures it is grouped with other decays with a similar strong-phase difference, which maximises the sensitivity to $\gamma$. Bin $i$ is then divided in two, with labels $i$ and $-i$ according to the value of $r_{\D}$, giving $2 \times {\mathcal N}$ bins in total. This division is performed in a manner to enhance the difference between $F_{+i}$ and $F_{-i}$, which maximises the magnitude of the interference terms in Eqs.~\eqref{equation:Ni_plus} and \eqref{equation:Ni_minus}. Candidates with $\ln r_\D < 0$ ($> 0$) are assigned to bin $i > 0$ ($< 0$), and the bin numbering starts at $i = \mathcal{N}$ ($i = 1$) near $\Delta\delta_\D = -\pi$, with decreasing (increasing) bin numbers.

Following the procedure described in Ref.~\cite{HarnewS4pi}, the binning scheme is optimised by adjusting the bin boundaries in $\Delta \delta_D$ to maximise the $Q$-value, 
 defined by \mbox{$Q^2 \equiv (Q_+^2 + Q_-^2)/2$}, where
\begin{equation}
    Q_\pm^2 \equiv 1 - \Big(1 - \sum_i\frac{F_iF_{-i}\big(1 - c_i^2 - s_i^2\big)}{N_{i}^\pm}\Big)\Big(\sum_iF_i\Big)^{-1}.
    \label{equation:Qvalue}
\end{equation}
Here $N_i^\pm$, the $B^\pm \to DK^\pm$ decay yield in bin $i$, as predicted by Eqs.~\eqref{equation:Ni_plus} and~\eqref{equation:Ni_minus}, is calculated with the normalisation coefficients $h_{B^\pm}$ set to unity. The parameters $Q_{\pm}$ give the statistical sensitivity of $x_\pm$ and $y_\pm$ from a binned fit, divided by that of an unbinned fit. With an infinite number of bins, $Q\to 1$. This metric neglects any perturbation to the sensitivity that may arise from the presence of background events.

By allowing for only even values of $\mathcal{N}$, the bin boundaries at $\Delta\delta_\D = 0$ and $\pm\pi$ may be fixed. Then each pair of bin boundaries on either side of $\Delta\delta_\D = 0$ are simultaneously adjusted until the $Q$-value is maximal. In each iteration $F_i$, $c_i$ and $s_i$  are calculated from the amplitude model using Eqs.~\eqref{equation:Fi} and~\eqref{equation:ci}, assuming a uniform acceptance. The five-dimensional integral over phase space is performed with Monte Carlo integration. Large samples of $\D$-decays are generated for the integration, such that the uncertainty due to the finite sample size is negligible. The values of $N_i^\pm$ are determined from Eqs.~\eqref{equation:Ni_plus} and \eqref{equation:Ni_minus}, and take as input $\gamma = 75^\circ$, $\delta^{DK}_\B = 130^\circ$ and $r^{DK}_B = 0.1$, which lie close to the known values of these parameters~\cite{LHCb-PAPER-2021-033}. Note, however, that $Q$ has a very weak dependence on the values of the phases assumed in these expressions, so this choice does not bias the analysis.

\begin{figure}[tb]
    \centering
    \begin{subfigure}{0.57\textwidth}
        \includegraphics[height=6.2cm]{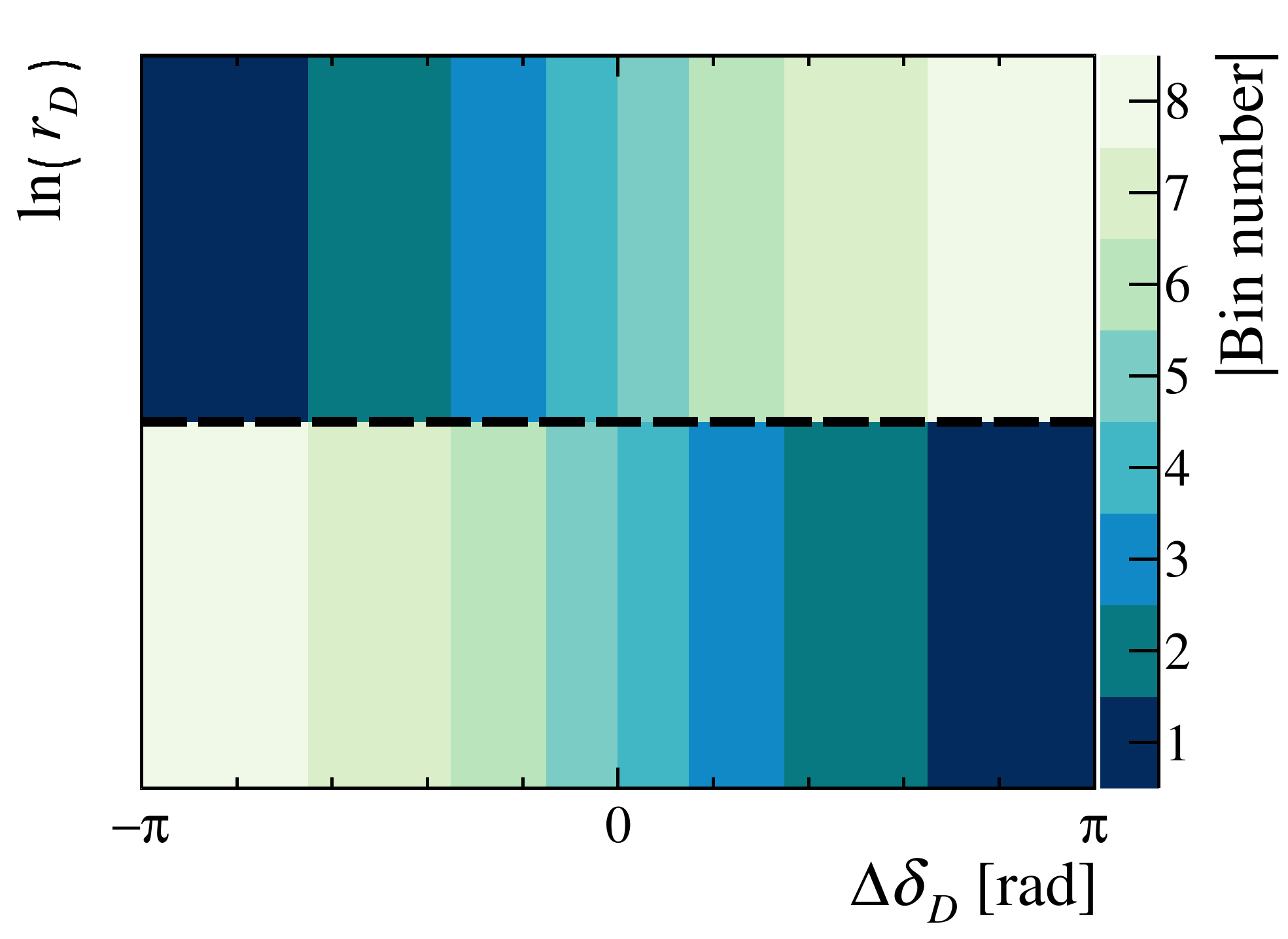}
    \end{subfigure}%
    \hfill
    \begin{subfigure}{0.43\textwidth}
        \includegraphics[height=6.2cm]{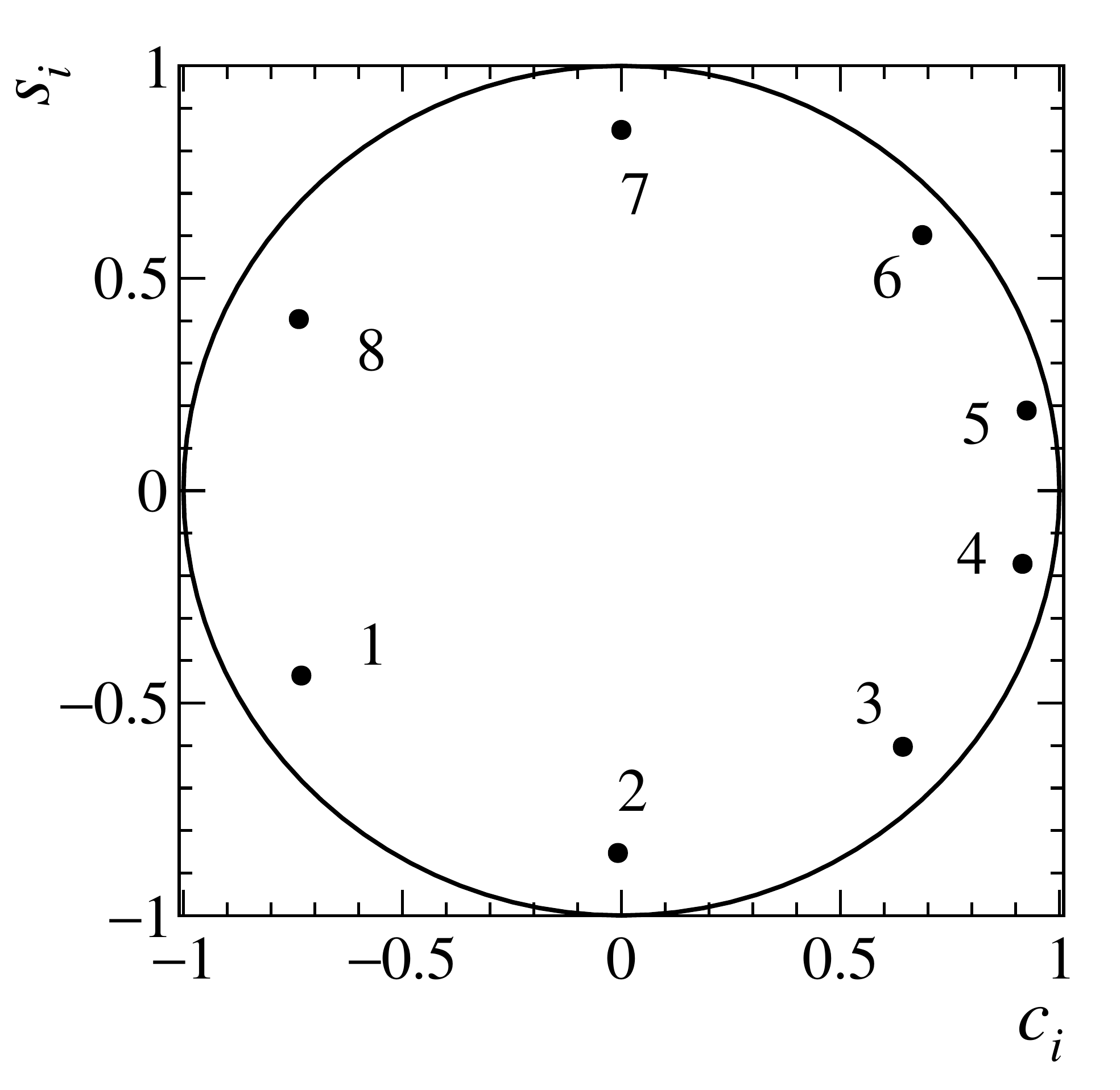}
    \end{subfigure}
    \caption{Left: Optimised $2\times 8$ binning scheme in $\Delta \delta_D$-$\ln(r_D)$ space. Right: The associated $c_i$ and $s_i$ parameters calculated using the amplitude model (right). The numbers indicate the bin numbers.}
    \label{figure:Binning_scheme_plots_8bins}
\end{figure}

Figure~\ref{figure:Binning_scheme_plots_8bins}, on the left, shows the binning scheme resulting from this procedure for $2 \times 8$ bins, and the values of $c_i$ and $s_i$ predicted by the amplitude model are shown in Table~\ref{table:ci_si_Fi_2x8}, along with the predicted values of $F_i$ and the fractional bin volume $V_i$. In the calculation, the region of phase space where the invariant mass of the $\pip\pim$ lies close to the $\KS$ mass is excluded. This requirement, which removes around $5\%$ of signal decays, is imposed to match a selection requirement to remove background that is described in Sec.~\ref{section:Candidate_selection}.

The binning scheme shown on the left in Fig.~\ref{figure:Binning_scheme_plots_8bins} has $Q = 0.90$, which indicates that only $10\%$ of the statistical sensitivity is lost through the binning of phase space. The procedure for assigning bin numbers for this scheme according to the four-momenta of the $\D$ decay is provided in Ref.~\cite{cite:KKpipiBinningScheme}. An alternative optimised binning scheme with $\mathcal{N}=4$ is presented in Appendix~\ref{section:Yields_in_bins_of_phase_space}.

The determination of the $\CP$-violating parameters requires as input the values of $c_i$ and $s_i$ in each bin. Although the bins have been defined using the amplitude model, it will be possible to use direct measurements of these parameters made at charm threshold, when they become available. The choice of binning scheme does not bias the determination of the $\CP$-violating parameters, even if the direct measurements were to indicate that the amplitude model gives an imperfect description of the strong-phase variation.  In this case, there would be a reduction in statistical sensitivity compared to current expectations, but the result obtained for $\gamma$ would have no model-dependent uncertainty.  

\begin{table}[tb]
    \centering
    \caption{Values of $c_i$, $s_i$, $F_i$ and $V_i$ for the optimised $2\times 8$ binning scheme, as calculated from the amplitude model. The number of digits quoted reflects the uncertainty due to the sample size used in the Monte Carlo integration.}
    \label{table:ci_si_Fi_2x8}
    \begin{tabular}{cccccccc}
        \toprule
        Bin number& $c_i$     & $s_i$     & $F_i$     & $F_{-i}$  & $V_i$     & $V_{-i}$ \\
        \midrule
        $1$       & $-0.7317$ & $-0.4343$ & $0.0157$  & $0.0495$  & $0.0555$  & $0.0555$   \\
        $2$       & $-0.0076$ & $-0.8528$ & $0.0185$  & $0.0644$  & $0.0645$  & $0.0645$   \\
        $3$       & $\phantom{-}0.6406$  & $-0.6056$ & $0.0295$  & $0.1024$  & $0.0753$  & $0.0754$   \\
        $4$       & $\phantom{-}0.9151$  & $-0.1728$ & $0.0687$  & $0.1466$  & $0.0654$  & $0.0655$   \\
        $5$       & $\phantom{-}0.9247$  & $\phantom{-}0.1887$  & $0.0815$  & $0.1646$  & $0.0742$  & $0.0742$   \\
        $6$       & $\phantom{-}0.6853$  & $\phantom{-}0.6021$  & $0.0398$  & $0.0973$  & $0.0665$  & $0.0664$   \\
        $7$       & $-0.0032$ & $\phantom{-}0.8490$  & $0.0143$  & $0.0488$  & $0.0510$  & $0.0510$   \\
        $8$       & $-0.7368$ & $\phantom{-}0.4041$  & $0.0132$  & $0.0451$  & $0.0475$  & $0.0476$   \\
        \bottomrule
    \end{tabular}
\end{table}


\section{The LHCb detector and data set}
\label{section:The_LHCb_detector_and_data_set}
This analysis uses data collected by the \lhcb experiment in $pp$ collisions at $\sqrt{s} = 7\tev$, $8\tev$ and $13\tev$. The data sets correspond to integrated luminosities of $1\invfb$, $2\invfb$ and $6\invfb$, respectively.

The \lhcb detector~\cite{LHCb-DP-2008-001,LHCb-DP-2014-002} is a single-arm forward spectrometer covering the \mbox{pseudorapidity} range $2<\eta <5$, designed for the study of particles containing \bquark or \cquark quarks. The detector includes a high-precision tracking system consisting of a silicon-strip vertex detector surrounding the $pp$ interaction region, a large-area silicon-strip detector located upstream of a dipole magnet with a bending power of about $4{\mathrm{\,Tm}}$, and three stations of silicon-strip detectors and straw drift tubes placed downstream of the magnet. The tracking system provides a measurement of the momentum, \ptot, of charged particles with a relative uncertainty that varies from 0.5\% at low momentum to 1.0\% at 200\gevc. The minimum distance of a track to a primary $pp$ collision vertex (PV), the impact parameter (IP), is measured with a resolution of $(15+29/\pt)\mum$, where \pt is the component of the momentum transverse to the beam, in\,\gevc. Different types of charged hadrons are distinguished using information from two ring-imaging Cherenkov detectors. Photons, electrons and hadrons are identified by a calorimeter system consisting of scintillating-pad and preshower detectors, an electromagnetic and a hadronic calorimeter. Muons are identified by a system composed of alternating layers of iron and multiwire proportional chambers. The online event selection is performed by a trigger, which consists of a hardware stage, based on information from the calorimeter and muon systems, followed by a software stage, which applies a full event reconstruction.

Simulation is required to model the effects of the detector acceptance, and for the study of possible background processes. In the simulation, $pp$ collisions are generated using \pythia~\cite{Sjostrand:2007gs,*Sjostrand:2006za} with a specific \lhcb configuration~\cite{LHCb-PROC-2010-056}. Decays of unstable particles are described by \evtgen~\cite{Lange:2001uf}, in which final-state radiation is generated using \photos~\cite{davidson2015photos}. The $\Dz\to\Kp\Km\pip\pim$ decay is simulated using the amplitude model from Ref.~\cite{LHCb-PAPER-2018-041}. The interaction of the generated particles with the detector, and its response, are implemented using the \geant toolkit~\cite{Allison:2006ve, *Agostinelli:2002hh} as described in Ref.~\cite{LHCb-PROC-2011-006}. The underlying $pp$ interaction is reused multiple times, with an independently generated signal decay for each event~\cite{LHCb-DP-2018-004}.


\section{Candidate selection}
\label{section:Candidate_selection}
A $\Bpm$ candidate is reconstructed by combining five charged tracks. Four of the charged tracks are required to have an invariant mass within $25\mevcc$ of the $\Dz$ meson mass~\cite{PDG2022}. This requirement, which corresponds to around two and a half times the resolution of the mass peak, removes processes that have either a missing or misidentified particle. Candidates where the opening angle between any pair of tracks from the $\D$-decay products is smaller than $0.03^\circ$ are discarded, as these are likely to correspond to a single charged particle that is duplicated in the reconstruction.

To suppress charmless background, which arises from $\Bpm$ meson decays where there is no intermediate charm meson, the distance between the $\D$ and $\Bpm$ decay vertices is required to be greater than twice its resolution for the binned measurement. This criterion eliminates $95\%$ of this category of decays. In the phase-space integrated measurement, which is found to be more sensitive to this source of contamination, the decay distance requirement is tightened to four times its resolution, suppressing the background by a further order of magnitude.

Separation of $\Bpm\to\D\Kpm$ and $\Bpm\to\D\pipm$ decays is achieved by imposing mutually exclusive particle identification (PID) requirements on the companion track, which is the $\Kpm$ or $\pipm$ meson of the $\Bpm\to\D h^\pm$ decay. Companion tracks that have associated activity in the muon detector are removed; this requirement reduces background from semi-leptonic $b$-hadron decays involving a muon which is misidentified as a companion kaon or pion. Background from semi-leptonic $b$-decays involving an electron is found to be negligible.  

Background from $\D$ decays where a $\pip\pim$ pair originates from a $\KS$ meson is suppressed by excluding regions containing $\pip\pim$ pairs with invariant mass inside the interval $[477, 507]\mevcc$ from the binning scheme. Additionally, PID requirements are imposed on the kaon from the $\D$ candidate with opposite sign to the companion track. This selection requirement suppresses $\D\to\Kmp\pipm\pim\pip\piz$ background where the $\piz$ meson is not reconstructed and the kaon is misidentified. 

Combinatorial background is suppressed using a boosted decision tree~(BDT) algorithm~\cite{Breiman,AdaBoost} implemented in the TMVA toolkit~\cite{Hocker:2007ht,*TMVA4}. Simulated signal events are used as the signal training sample, while candidates with invariant mass in the upper $\Bpm$ sideband between $5800$--$7000$\mevcc form the background training sample. The input variables of the BDT include the momenta and IPs of the $\Bpm$, $\D$ and companion-track candidates. The parameters are described in detail in Ref.~\cite{LHCb-PAPER-2018-017}. The optimal working point of the BDT is determined by performing pseudoexperiments to determine the cut value that provides the best sensitivity to $\gamma$. Candidates with a BDT score below this cut value are discarded.

To improve the resolution of the momenta of the $\D$-decay products and the invariant mass of the $\Bpm$ candidate, a kinematic fit is performed in which the $\D$ meson candidate is constrained to its known mass~\cite{PDG2022}, and the $\Bpm$ candidate is constrained to originate from its associated PV. This is defined as the PV with the smallest impact parameter with respect to the $\Bpm$ candidate.


\section{Invariant-mass fits}
\label{section:Invariant_mass_fits}

\begin{figure}[tb]
    \centering
    \includegraphics[width=1.0\textwidth]{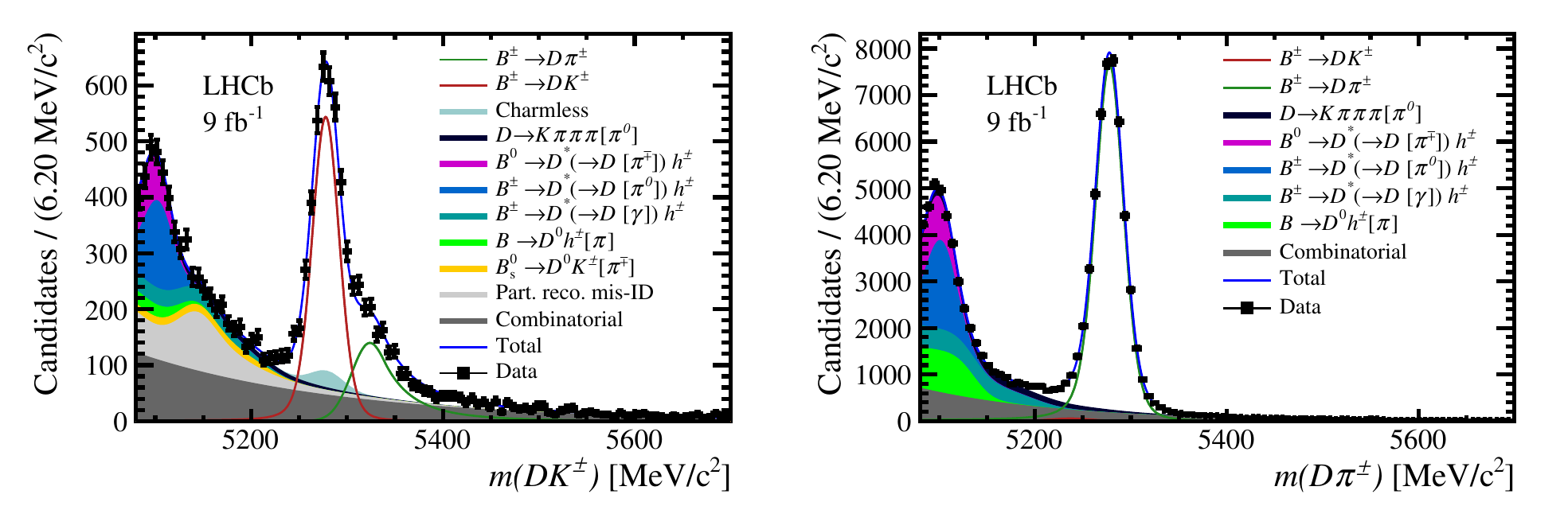}
    \caption{Invariant-mass distributions for the (left) $\Bpm\to\D\Kpm$ and (right) $\Bpm\to\D\pipm$ selections, for the $\D\to\Kp\Km\pip\pim$ decay. The data are shown as black points and the blue curve is the fit result. The square brackets in the legend denote particles that are not reconstructed.}
    \label{figure:Global_fit}
\end{figure}

An unbinned, extended maximum-likelihood fit is performed simultaneously to the invariant-mass spectrum of the $\Bpm\to[\Km\Kp\pip\pim]_\D h^\pm$ and $\Bpm\to[\pim\pip\pip\pim]_\D h^\pm$ candidates in the range from $5080\mevcc$ to $5700\mevcc$. The fit is first performed on the $\Bpm\to\D\Kpm$ and $\Bpm\to\D\pipm$ candidates, integrated over all phase-space bins, which is referred to as the global fit. The global fit is used to determine the parameters of the functions that describe the signal and background invariant-mass distributions. The $\Bpm\to[\Km\Kp\pip\pim]_\D h^\pm$ invariant-mass distributions are shown in Fig.~\ref{figure:Global_fit}.

The yield of $\Bpm\to\D\pipm$ is varied separately for the two $\D$ decays, while the yield of $\Bpm\to\D\Kpm$ is parameterised as a ratio relative to the $\Bpm\to\D\pipm$ yield. This ratio is a common fit parameter for the two $\D$ decays, as are the parameters that describe the signal shape.

The peak at around $5280\mevcc$ corresponds to $\Bpm\to[\Kp\Km\pip\pim]_\D h^\pm$ candidates that are correctly reconstructed. The signal invariant-mass shape is parameterised as
\begin{equation}
    \begin{aligned}
        f_{\rm signal}(m|m_\B, \sigma, \alpha_L, \alpha_R, \beta, k) =& k\times f_{\rm MG}(m|m_\B, \sigma, \alpha_L, \alpha_R, \beta) \\
                                                                     +& (1 - k)\times f_{\rm G}(m|m_\B, \sigma),
    \end{aligned}
\end{equation}
where $f_{\rm G}$ is a Gaussian function and $f_{\rm MG}$ is a modified Gaussian function,

\begin{equation}
    f_{\rm MG}(m|m_\B, \sigma, \alpha_L, \alpha_R, \beta)\propto
    \begin{cases}
        \exp\Big(-\frac{\Delta m^2(1 + \beta\Delta m^2)}{2\sigma^2 + \alpha_L\Delta m^2}\Big), \quad \Delta m = m - m_\B < 0, \\
        \exp\Big(-\frac{\Delta m^2(1 + \beta\Delta m^2)}{2\sigma^2 + \alpha_R\Delta m^2}\Big), \quad \Delta m = m - m_\B > 0.
    \end{cases}
\end{equation}
The function $f_{\rm MG}$ has approximately Gaussian behaviour when $\Delta m^2\ll\sigma^2/\alpha_{L, R}$ or $\Delta m^2\gg\beta^{-1}$, but it includes tails to better model the experimental resolution. The tail parameters $\alpha_{L, R}$ and $\beta$, and the fraction $k$, are determined in a fit to simulated events, but the peak position $m_\B$ and the width $\sigma$ are determined in the fit to data. The mass $m_\B$ is common between the $\Bpm\to\D\Kpm$ and $\Bpm\to\D\pipm$ channels, but $\sigma$ is different because the $\Bpm\to\D\Kpm$ width is narrower due to the lower energy release in the decay.

At masses above the $\Bpm\to\D\Kpm$ peak there is a non-negligible contribution from $\Bpm\to\D\pipm$ decays where the companion is misidentified as a kaon. The rate of this cross-feed background is fixed from the relative PID efficiencies, which are determined in calibration data that are weighted to match the momentum and pseudorapidity distributions of the companion track of the signal. The exact shape is determined using a data-driven method by swapping the mass hypothesis of the companion track in the $\Bpm\to\D\pipm$ peak. Similarly, the shape of $\Bpm\to\D\Kpm$ candidates misidentified as $\Bpm\to\D\pipm$ candidates is also accounted for, but the impact of this background is minimal due to its smaller branching fraction.

Candidates with masses below that of the signal peak are background from $\B$-meson decays where a neutral particle or charged pion is not reconstructed. In this analysis, the model describing this partially reconstructed background and its associated parameters are taken from Ref.~\cite{LHCb-PAPER-2020-019}, with the exception of the contamination from $\Bs\to\Dzb\Km\pip$ and $\Bsb\to\Dz\Kp\pim$ decays with a missing pion. The total yield of this latter background, which is fixed in the fit, has been updated according to results in Ref.~\cite{LHCb-PAPER-2020-036}.

Additionally, the decay $\D\to\Kp\Km\pip\pim$ is contaminated by $\D\to\Kmp\pipm\pim\pip\piz$ decays, where a charged pion is misidentified as a kaon and the neutral pion is not reconstructed. This background is present below the signal peak, but it has a large tail towards the upper end of the invariant mass spectrum of the $\Bpm$ candidates as well. The shape of this background is fixed using a simulation sample, while its yield is a free parameter. The ratio between this background and signal is common between the $\Bpm\to\D\Kpm$ and $\Bpm\to\D\pipm$ modes.

The contamination of charmless decays in the $\D\to\Kp\Km\pip\pim$ mode is also different from Ref.~\cite{LHCb-PAPER-2020-019}. In particular, the $\Bpm\to[\Kp\Km\pip\pim]_\D\Kpm$ sample has a significant contribution from the mode $\Bpm\to\Kp\Km\pip\pim\Kpm$. Both the magnitude and invariant-mass shape of this contribution are fixed in the invariant-mass fit from studies of the lower sideband of the $\D$ invariant mass. Analogous studies show that there is no significant contamination from charmless decays in the $\Bpm\to\D\pipm$ selection.

The signal yields, obtained from the invariant-mass fit, integrated over all phase-space bins, are given in Table~\ref{table:Global_fit_yields}. The yields are scaled from the full fit region to the signal region $m_\B\in[5249, 5309]\mevcc$. The uncertainties on the $\Bpm\to\D\Kpm$ yields are reduced due to the common ratio determined from both the $\D\to\Kp\Km\pip\pim$ and $\D\to\pip\pim\pip\pim$ decay modes.

\begin{table}[tb]
    \centering
    \caption{Yields of $\Bpm\to\D\Kpm$ and $\Bpm\to\D\pipm$ candidates, partially reconstructed background, $\D\to\Km\pip\pim\pip\piz$ background, combinatorial background and charmless background in the region $m_\B\in[5249, 5309]\mevcc$, where the charm meson decays via $\D\to\Kp\Km\pip\pim$ and $\D\to\pip\pim\pip\pim$. Charmless background in the $\Bpm\to\D\pipm$ decay modes is known to be negligible, and is therefore not modelled.}
    \label{table:Global_fit_yields}
    \begin{tabular}{lcccc} 
        \toprule
                                 &                          & \multicolumn{2}{c}{Reconstructed as:} \\
        $\D$ decay               & Component                & $\Bpm\to\D\Kpm$          & $\Bpm\to\D\pipm$         \\
        \midrule
        \D\to~\Kp\Km\pip\pim     & \Bpm\to\D\Kpm            & $3026 \pm 38\phantom{0}$            & $\phantom{000}142 \pm 2\phantom{00}$              \\
                                 & \Bpm\to\D\pipm           & $\phantom{0}240 \pm 1\phantom{00}$              & $\phantom{0}44349 \pm 218$          \\
                                 & Partially reconstructed bkg & $\phantom{00}87 \pm 1\phantom{00}$               & $\phantom{0000}27 \pm 1\phantom{00}$               \\
                                 & \D\to~\Kmp\pipm\pim\pip\piz & $\phantom{00}44 \pm 13\phantom{0}$              & $\phantom{000}580 \pm 168$            \\
                                 & Combinatorial bkg & $\phantom{0}460 \pm 23\phantom{0}$             & $\phantom{00}1820 \pm 193$           \\
                                 & Charmless bkg            & \phantom{00}$189$ (fixed)              & Not modelled                \\
        \midrule
        \D\to~\pip\pim\pip\pim   & \Bpm\to\D\Kpm            & $8676 \pm 105$           & $\phantom{000}386 \pm 5\phantom{00}$              \\
                                 & \Bpm\to\D\pipm           & $\phantom{0}676 \pm 2\phantom{00}$              & $126322 \pm 386$         \\
                                 & Partially reconstructed bkg   & $\phantom{0}256 \pm 2\phantom{00}$              & $\phantom{0000}81 \pm 4\phantom{00}$               \\
                                 & Combinatorial bkg & $1344 \pm 27\phantom{0}$            & $\phantom{00}4172 \pm 90\phantom{0}$            \\
                                 & Charmless bkg            & \phantom{00}$688$ (fixed)              & Not modelled                \\
        \bottomrule
    \end{tabular}
\end{table}

After the global invariant-mass fit, a second fit is performed where the $\Bpm\to\D\Kpm$ and $\Bpm\to\D\pipm$ candidates are split by charge and sorted into bins of phase space, which makes a total of $2\times 2\times 16 = 64$ categories. The lower fit boundary is increased to $5150\mevcc$ to remove most of the partially reconstructed background. The shape parameters and relative yields of the different background components are fixed from the global fit. The signal yields in each bin are parameterised in terms of the \CP-violating observables, which are free parameters in the fit. The $F_i$ parameters are also free parameters, while the strong-phase parameters $c_i$ and $s_i$ are fixed according to the \lhcb amplitude model.

In each bin, the yield of combinatorial background and partially reconstructed background are free parameters, with the exception of the $\Bs\to\Dzb\Km\pip$ contamination (and charge-conjugated case), which is treated separately because the charm meson has the flavour opposite to the signal decay and the other partially reconstructed background contributions. In $\Bm$ ($\Bp$) decays, the fractional bin yield of the $\Bs$ (\Bsb) background is therefore set equal to $F_{-i}$ ($F_i$). In simulation, the $\D \to\Kmp\pipm\pim\pip\piz$ decays are uniformly distributed in the $\D \to\Kp\Km\pip\pim$ phase space. Therefore, the distribution of $\D \to\Kmp\pipm\pim\pip\piz$ decays between phase space bins is assumed to be proportional to the bin volume, given in Table~\ref{table:ci_si_Fi_2x8}. The distribution of the charmless background between phase-space bins is determined from the lower $\D$-mass sideband.

Fit biases and instabilities in the fit are studied by performing pseudoexperiments. The pull distributions of $x^{DK}_\pm$ and $y^{DK}_\pm$ are found to be consistent with a normal Gaussian distribution. The results for $x^{D\pi}_\xi$ and $y^{D\pi}_\xi$ show biases of up to $7\%$ and widths that show up to $17\%$ overcoverage. Corrections are applied to the measured values in the data to account for these effects.

\begin{figure}[tb]
    \centering
    \begin{subfigure}{0.5\textwidth}
        \includegraphics[width=1\textwidth]{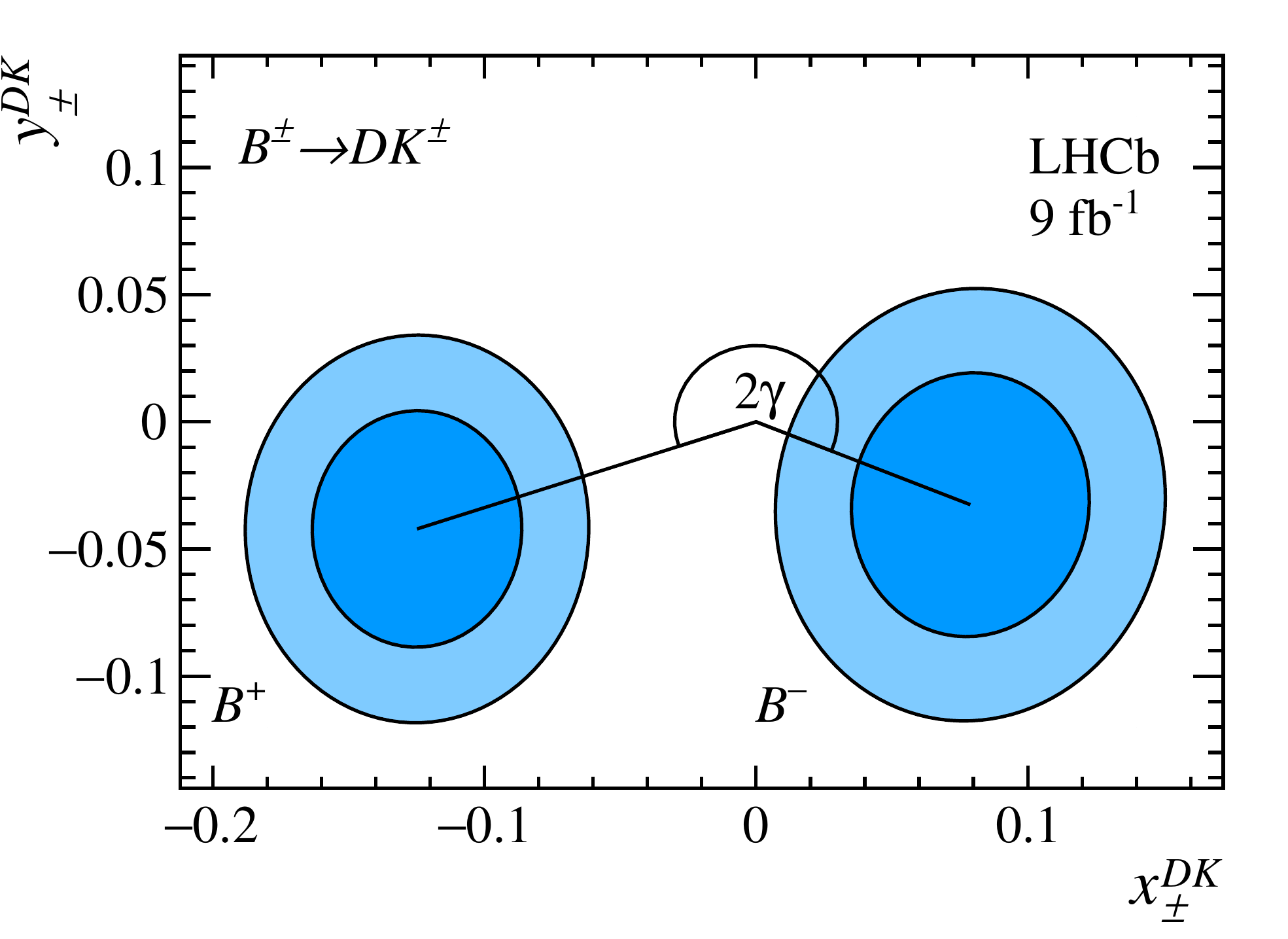}
    \end{subfigure}%
    \begin{subfigure}{0.5\textwidth}
        \includegraphics[width=1\textwidth]{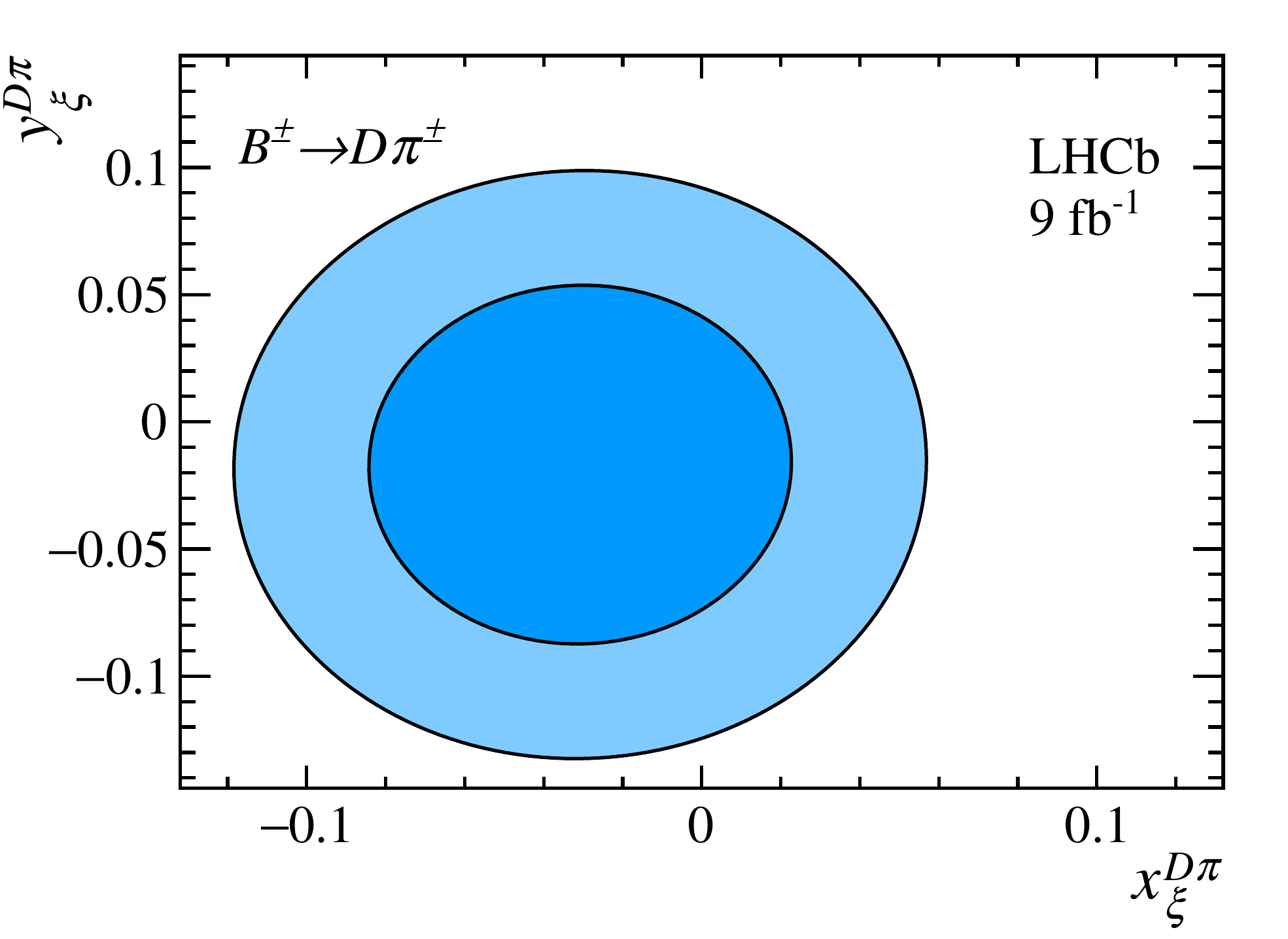}
    \end{subfigure}
    \caption{Graphical representation of the \CP-violating observables for the (left) $\Bpm\to\D\Kpm$ and (right) $\Bpm\to\D\pipm$ decay modes, assuming amplitude-model predictions of $c_i$ and $s_i$. The 1$\sigma$ and 2$\sigma$ contours shown represent the statistical uncertainties and correspond to $68.3\%$ ($\Delta\chi^2 = 2.30$) and $95.5\%$ ($\Delta\chi^2 = 6.18$) confidence intervals.}
    \label{figure:CP_observables}
\end{figure}

The fitted \CP-violating observables are listed in Table~\ref{table:GGSZ_observables} and plotted in Fig.~\ref{figure:CP_observables}, along with the likelihood contours, which only include statistical uncertainties. The lengths of the two vectors from the origin to $(x_\pm^{\D\kaon}, y_\pm^{\D\kaon})$ determine $r_\B^{\D\kaon}$, while the angle between them is $2\gamma$, according to the expressions in Eq.~\eqref{equation:CP_observables}. In the absence of \CP violation, the two vectors would be identical. The vector from the origin to $(x_\xi^{\D\pion}, y_\xi^{\D\pion})$ indicates the relative size and angle between $r_\B$ and $\delta_\B$ of the $\Bpm\to\D\pipm$ and $\Bpm\to\D\Kpm$ decays, according to Eq.~\eqref{equation:xi_observables}. Since this contour overlaps with the origin, the sensitivity to \CP violation is much smaller in the $\Bpm\to\D\pipm$ mode.

\begin{table}[tb]
    \centering
    \caption{Results of the binned fit. The first uncertainty is statistical, the second is systematic, and the third is associated with the model dependence of the strong-phase parameters.}
    \label{table:GGSZ_observables}
    \begin{tabular}{cc} 
        \toprule
        \CP-violating observable & Fit result ($\times 10^2$)           \\
        \midrule
        $x_-^{\D\kaon}$      & \phantom{+}$7.9 \pm 2.9 \pm 0.4 \pm 0.4$        \\
        $y_-^{\D\kaon}$      & $-3.3 \pm 3.4 \pm 0.4 \pm 3.6$       \\
        $x_+^{\D\kaon}$      & $-12.5 \pm 2.5 \pm 0.3 \pm 1.7$\phantom{0}      \\
        $y_+^{\D\kaon}$      & $-4.2 \pm 3.1 \pm 0.3 \pm 1.3$ \\
        $x_\xi^{\D\pion}$    & $-3.1 \pm 3.5 \pm 0.7 \pm 0.1$       \\
        $y_\xi^{\D\pion}$    & $-1.7 \pm 4.7 \pm 0.6 \pm 1.1$       \\
        \bottomrule
    \end{tabular}
\end{table}

\begin{figure}[tb]
    \centering
    \begin{subfigure}{0.5\textwidth}
        \includegraphics[width=1\textwidth]{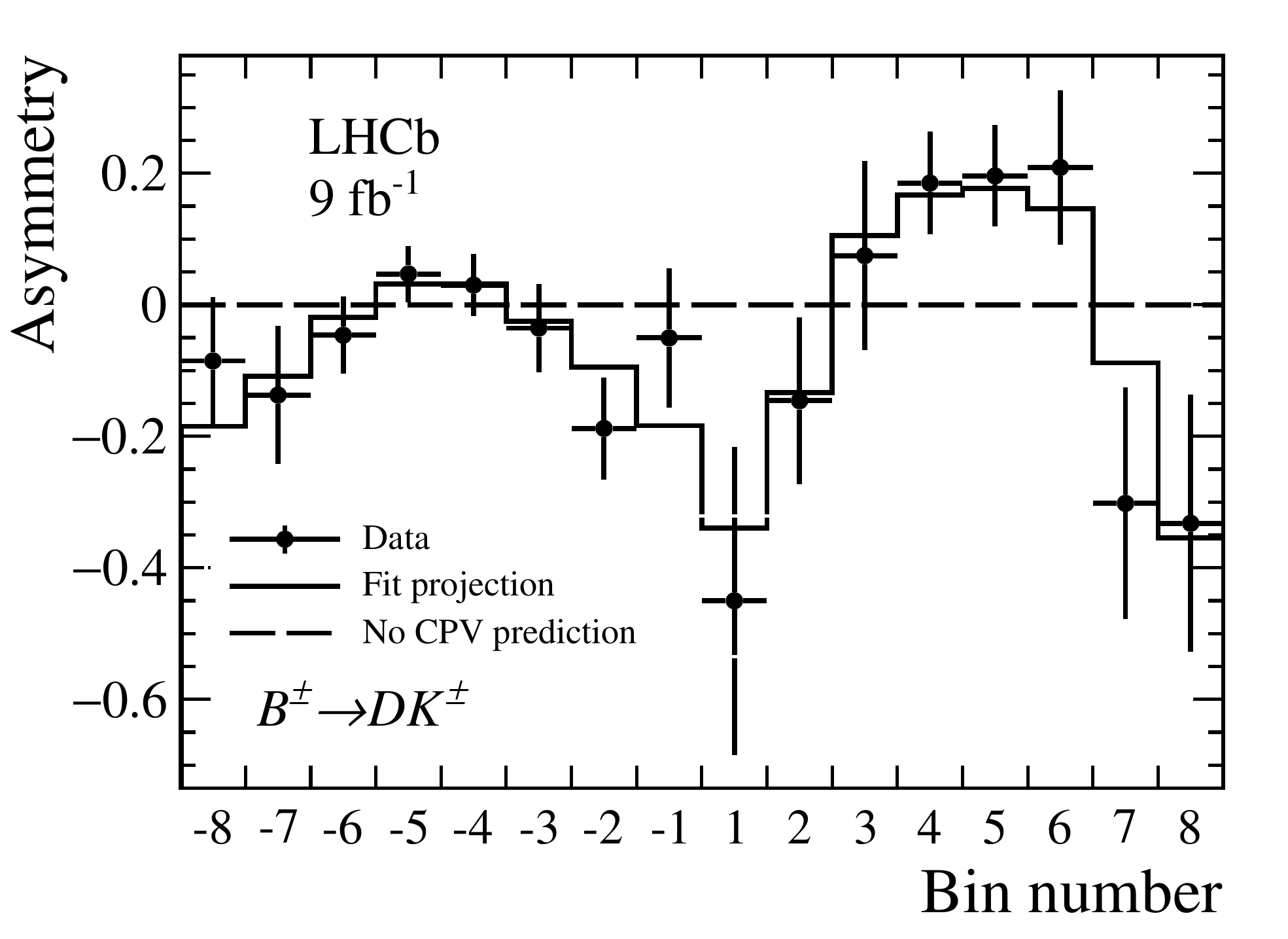}
    \end{subfigure}%
    \begin{subfigure}{0.5\textwidth}
        \includegraphics[width=1\textwidth]{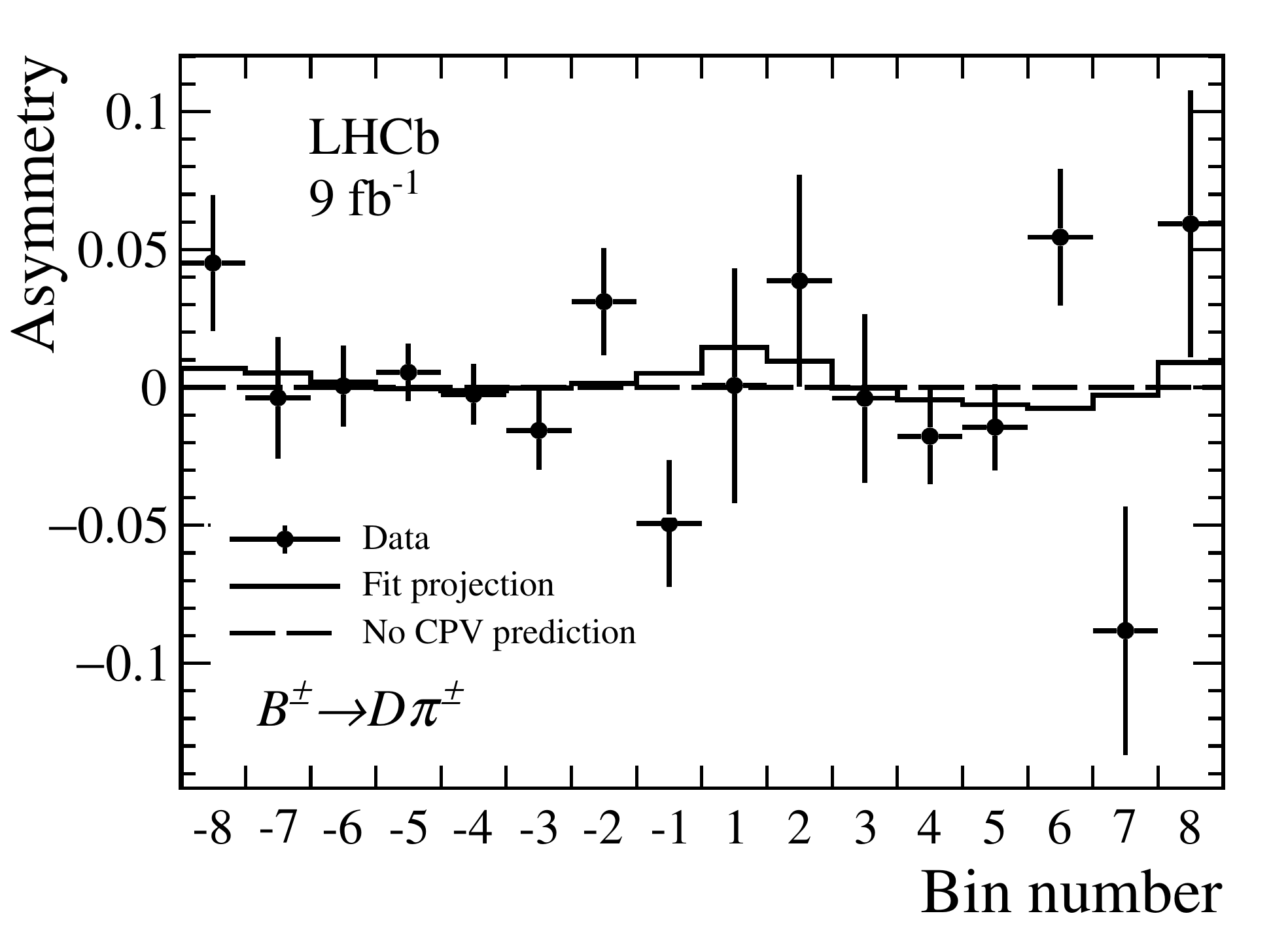}
    \end{subfigure}
    \caption{Fractional bin asymmetries for the (left) $\Bpm\to\D\Kpm$ and (right) $\Bpm\to\D\pipm$ decays. The data are overlaid with the fit result and the prediction without \CP violation.}
    \label{figure:Bin_asymmetries}
\end{figure}

The \CP-violation effects can be illustrated directly by considering the asymmetries in each bin. This information is obtained from an alternative fit where the signal yields, instead of the \CP-violating observables, are determined. The bin asymmetries, $(N_i^- - N_{-i}^+)/(N_i^- + N_{-i}^+)$, calculated from these yields are shown in Fig.~\ref{figure:Bin_asymmetries}. The bin yields are normalised separately for $\Bm$ and $\Bp$, so that only bin-to-bin variations are apparent. Some of those for $\Bpm\to\D\Kpm$ are significant, and exhibit a non-trivial distribution, which is driven by the variation in the strong-phase difference between $\Dz$ and $\Dzb$ decays. In the $\Bpm\to\D\pipm$ mode, there is a lower sensitivity to \CP violation and therefore no such behaviour is seen. The hypothesis that the fit model is correct leads to p-values of $0.95$ and $0.05$ for the left and right histograms in Fig.~\ref{figure:Bin_asymmetries} respectively, based on statistical uncertainties only.

\begin{figure}[tb]
    \centering
    \begin{subfigure}{0.5\textwidth}
        \includegraphics[width=1\textwidth]{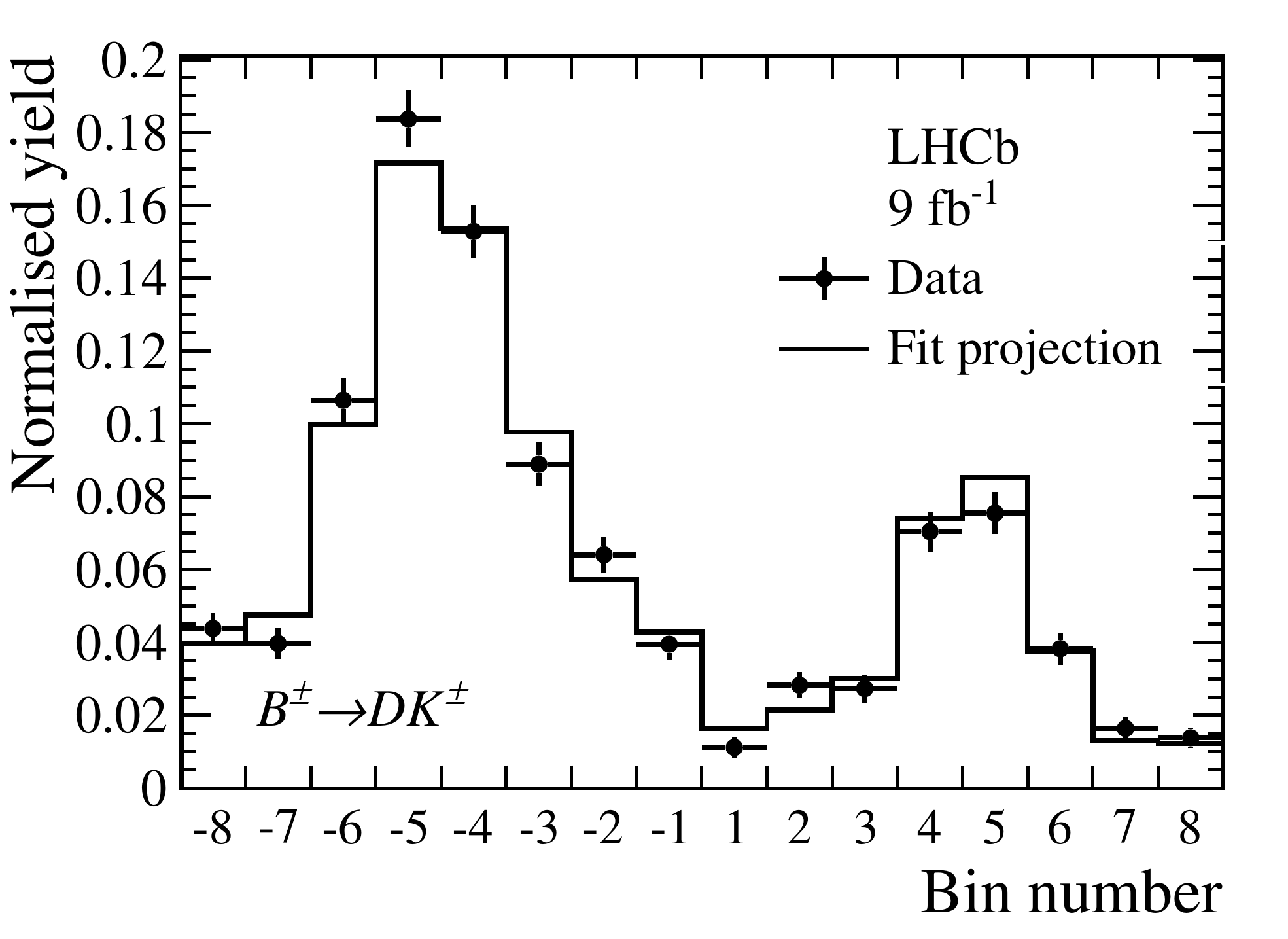}
    \end{subfigure}%
    \begin{subfigure}{0.5\textwidth}
        \includegraphics[width=1\textwidth]{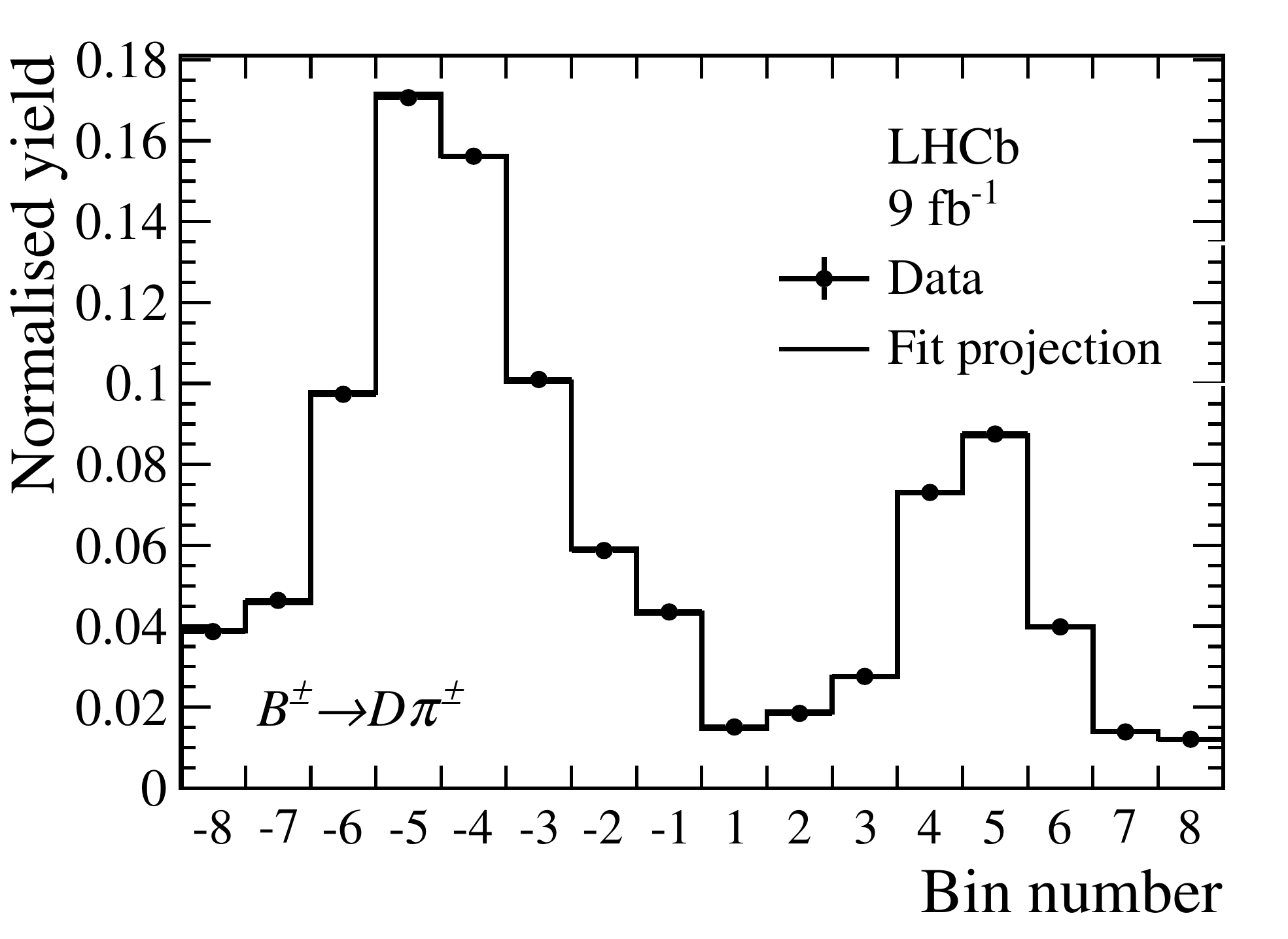}
    \end{subfigure}
    \caption{Total bin yields for the (left) $\Bpm\to\D\Kpm$ and (right) $\Bpm\to\D\pipm$ decays. The data are overlaid with the fit projections.}
    \label{figure:Total_bin_yields}
\end{figure}

The total bin yields $N_i^- + N_{-i}^+$ are also shown in Fig.~\ref{figure:Total_bin_yields}, where the sum over all bins is  normalised to unity. The projections from the fit results are also plotted, and reasonable agreement is found. Pseudoexperiments indicate that the p-values of the two histograms in Fig.~\ref{figure:Total_bin_yields} are highly correlated, and their combined p-value is $0.04$, accounting for statistical uncertainties only.

\begin{figure}[tb]
    \centering
    \includegraphics[width=1\textwidth]{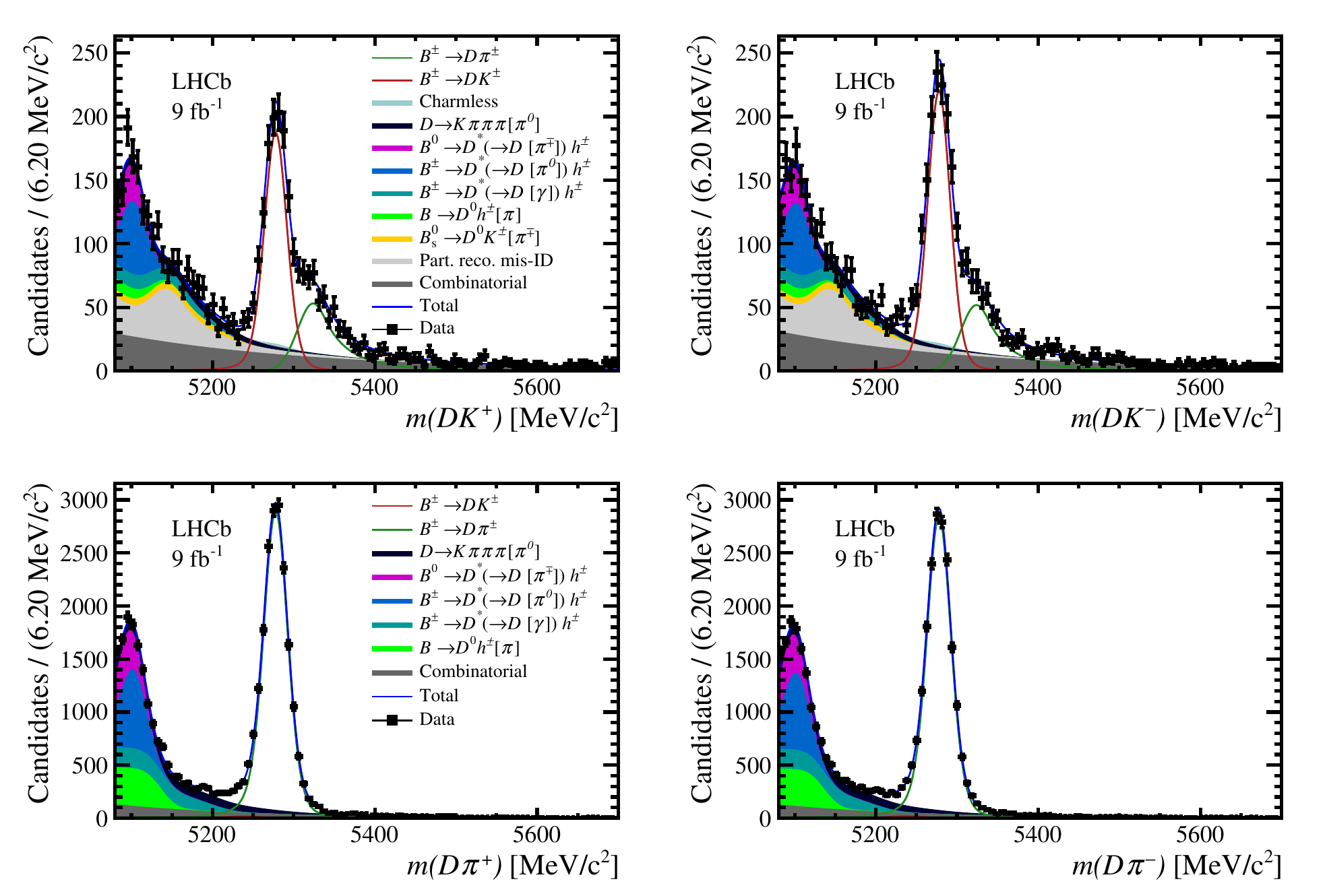}
    \caption{Invariant-mass distributions and fit projections of (top) $\Bpm\to[\Kp\Km\pip\pim]_\D\Kpm$ and (bottom) $\Bpm\to[\Kp\Km\pip\pim]_\D\pipm$ candidates, for (left) $\Bp$ and (right) $\Bm$ decays. The data are shown as black points and the blue curve is the fit result.}
    \label{figure:CP_observables_GLW_KKpipi}
\end{figure}

\begin{figure}[tb]
    \centering
    \includegraphics[width=1\textwidth]{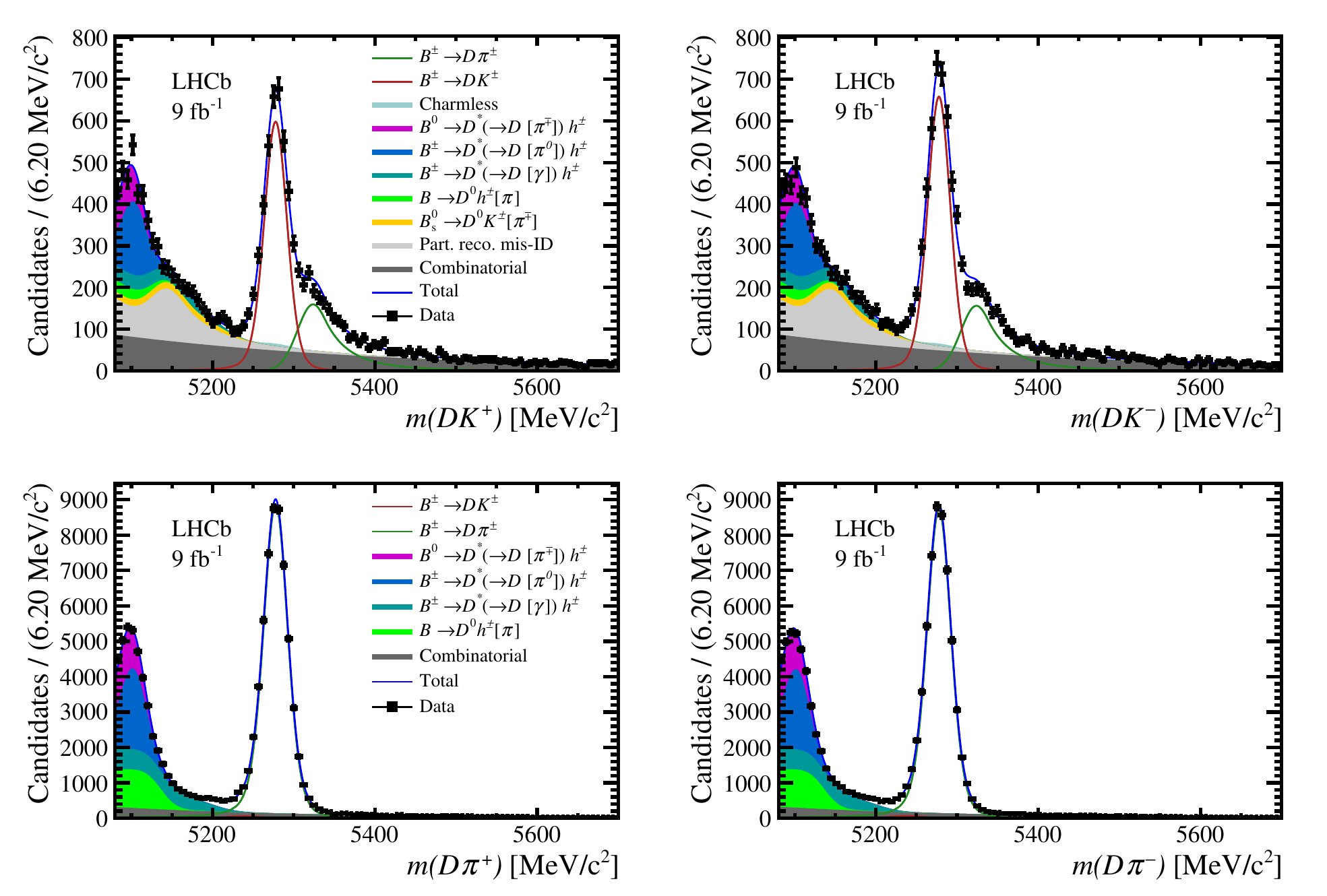}
    \caption{Invariant-mass  distributions and fit projections of (top) $\Bpm\to[\pip\pim\pip\pim]_\D\Kpm$ and (bottom) $\Bpm\to[\pip\pim\pip\pim]_\D \pipm$ candidates, for (left) $\Bp$ and (right) $\Bm$ decays. The data are shown as black points and the blue curve is the fit result.}
    \label{figure:CP_observables_GLW_pipipipi}
\end{figure}

To determine the phase-space integrated \CP-violating observables defined in Eqs.~\eqref{equation:Asymmetry} and ~\eqref{equation:R_CP}, an analogous fit is performed without phase-space binning. The yields of the $\Bpm\to[\Kp\Km\pip\pim]_\D h^\pm$ and $\Bpm\to[\pip\pim\pip\pim]_\D h^\pm$ modes, split by charge, are expressed in terms of the \CP-violating observables and fitted simultaneously. The shape parameters of the signal and background contributions are common fit parameters between the two $D$-decay channels. The fits to the invariant-mass distributions for the $\Bpm$ candidates are shown in Figs.~\ref{figure:CP_observables_GLW_KKpipi} and \ref{figure:CP_observables_GLW_pipipipi}, split by $\B$ decay, $\D$ decay and charge, and the resulting \CP-violating observables are listed in Table~\ref{table:GLW_observables}. The measured values of the observables of the $\Bpm\to[\pip\pim\pip\pim]_\D h^\pm$ mode are consistent with those reported in  Ref.~\cite{LHCb-PAPER-2016-003}.

The results in Table~\ref{table:GLW_observables} are corrected for production asymmetries of the $\Bpm$ mesons and detection asymmetries of the companion hadron. The production asymmetry was measured in Ref.~\cite{LHCb-PAPER-2020-036} to be $(0.028 \pm 0.068)\%$. In Ref.~\cite{LHCb-PAPER-2016-054}, the difference in detection asymmetries between kaons and pions was determined to be $(-0.96 \pm 0.13)\%$, in addition to a hardware-trigger asymmetry of $(-0.17 \pm 0.08)\%$. The detection asymmetry of pions was found to be $(0.064 \pm 0.018)\%$~\cite{LHCb-PAPER-2016-054}.

\begin{table}[tb]
    \centering
    \caption{Results of the phase-space integrated measurements. The first uncertainty is statistical and the second systematic.}
    \label{table:GLW_observables}
    \begin{tabular}{cc} 
        \toprule
        \CP-violating observable            & Fit results                            \\
        \midrule
        $A_{\kaon}^{\kaon\kaon\pion\pion}$  & $\phantom{+}0.095\phantom{0} \pm 0.023\phantom{0} \pm 0.002\phantom{0}$                   \\
        $A_{\pion}^{\kaon\kaon\pion\pion}$  & $-0.009\phantom{0} \pm 0.006\phantom{0} \pm 0.001\phantom{0}$                  \\
        $A_{\kaon}^{\pion\pion\pion\pion}$  & $\phantom{+}0.061\phantom{0} \pm 0.013\phantom{0} \pm 0.002\phantom{0}$                   \\
        $A_{\pion}^{\pion\pion\pion\pion}$  & $-0.0082 \pm 0.0031 \pm 0.0007$                \\
        $R_{\CP}^{\kaon\kaon\pion\pion}$ & $\phantom{+}0.974\phantom{0} \pm 0.024\phantom{0} \pm 0.015\phantom{0}$                   \\
        $R_{\CP}^{\pion\pion\pion\pion}$ & $\phantom{+}0.978\phantom{0} \pm 0.014\phantom{0} \pm 0.010\phantom{0}$                   \\
        \bottomrule
    \end{tabular}
\end{table}

\section{Systematic uncertainties}
\label{section:Systematic uncertainties}
The systematic uncertainties in the binned measurement are summarised in Table~\ref{table:Systematic_uncertainties}. The uncertainties arise both from contributions that are internal to the analysis, and also from external knowledge of the $c_i$ and $s_i$ parameters.

The uncertainty associated with the fixed invariant-mass shapes are propagated to the \CP-violating observables by repeating the two-stage fit procedure with different choices of shape. For each iteration, the shape parameters that are fixed in the global fit are changed to new values obtained with a resampling technique and the global fit is rerun. The other shape parameters that are determined from this fit are then input to the binned fit, which is otherwise unchanged from the baseline configuration. The standard deviations of the resulting distributions of the \CP-violating observables are assigned as the systematic uncertainty due to fixed mass shapes. Furthermore, to assess the impact of any bin-dependence of the mass shapes, the mass shapes are determined separately in each bin and pseudoexperiments are generated with individual mass shapes in each bin. The shifts in the central values are taken as the corresponding systematic uncertainty.

The uncertainties on the PID efficiencies are propagated to $x^{\D\kaon}_\pm$, $y^{\D\kaon}_\pm$, $x^{\D\pion}_\xi$ and $y^{\D\pion}_\xi$ by repeating the fit to the \CP-violating observables, each time varying the parameters within their uncertainties. The same procedure is followed to assign the uncertainty associated with the relative contributions of the different components of the low-mass partially reconstructed background. Similarly, for the charmless background, the yields are varied within their uncertainties. The standard deviations of the fitted \CP-violating observables are taken as the systematic uncertainty.

The partially reconstructed background at low mass is subject to \CP violation that means the bin distribution of these events differs between the $\Bp$ and $\Bm$ samples.  To investigate this effect, pseudoexperiments are generated containing \CP violation for these decays following the procedure described in Ref.~\cite{LHCb-PAPER-2020-019}, which are then fitted using the baseline model. The observed shifts in the central values are taken as the systematic uncertainty from this source.

There are several known sources of background present in the signal region that are not accounted for in the invariant-mass fit.  These are semi-leptonic $b$-hadron decays that survive the muon veto; $\Bpm\to\D h^\pm$ decays where the $\D$ meson decays semi-leptonically; $\Bpm\to\D h^\pm$, $\D\to\Kmp\pipm\pip\pim$ decays, where the kaon is misidentified as a pion, and two of the pions are misidentified as kaons; and decays of $\Lb\to\proton\Dz\pim$ (and charge conjugated case), where the proton is misidentified as the companion hadron and the pion is not reconstructed. To evaluate the potential bias arising from these neglected contributions, pseudoexperiments are generated with each component included, which are then fitted with the baseline model. The shifts in the resulting \CP-violating observables are taken as the systematic uncertainty. 

The distribution of the $\D\to\kaon^\mp\pion^\pm\pion^+\pion^-\piz$ background over phase space is not well known. The impact of this lack of knowledge is assessed by changing the distribution from that of the baseline model to one in which the population in each bin is proportional to the $F_i$ parameters. The shifts in the \CP-violating observables are assigned as the systematic uncertainty.

Finally, systematic uncertainty due to fit biases is included, which is set to be equal to the size of the bias for each \CP-violating observable.  Adding these in quadrature to the contributions discussed above gives a total internal systematic uncertainty for each observable, but excluding the systematic uncertainty arising from $c_i$ and $s_i$. The total internal systematic uncertainty is found to be an order of magnitude smaller than the corresponding statistical uncertainty.

In the current analysis the values of the $c_i$ and $s_i$ parameters are taken from the amplitude model constructed with \lhcb data and described in Ref.~\cite{LHCb-PAPER-2018-041}. An alternative model fitted to data from the \cleo experiment~\cite{CLEO_KKpipi_cisi} is used to generate pseudoexperiments which are then fitted using the $c_i$ and $s_i$ parameters from the \lhcb model. The observed shifts in the \CP-violating observables are taken as the uncertainties arising from the choice of model used to calculate the $c_i$ and $s_i$ parameters. These uncertainties are in several cases significantly larger than the \lhcb systematic uncertainties. In the future, the values of $c_i$ and $s_i$ will be taken from measurements performed at charm threshold, which will affect both the central values of the observables and allow the corresponding uncertainties to be assigned in a model-independent manner.

\begin{table}[tb]
    \centering
    \caption{Uncertainties on the results of the binned analysis.}
    \label{table:Systematic_uncertainties}
    \begin{tabular}{lcccccc}
        \toprule
        & \multicolumn{6}{c}{Uncertainty ($\times 10^2$)} \\
        \midrule
        Source & $x_-^{\D\kaon}$ & $y_-^{\D\kaon}$ & $x_+^{\D\kaon}$ & $y_+^{\D\kaon}$ & $x_\xi^{\D\pion}$ & $y_\xi^{\D\pion}$ \\
        \midrule
        Mass shape                                                 & $0.02$ & $0.02$ & $0.03$ & $0.06$ & $0.02$ & $0.04$ \\
        Bin-dependent mass shape                                   & $0.11$ & $0.05$ & $0.10$ & $0.19$ & $0.68$ & $0.16$ \\ 
        PID efficiency                                             & $0.02$ & $0.02$ & $0.03$ & $0.06$ & $0.02$ & $0.04$ \\
        Low-mass background model                                  & $0.02$ & $0.02$ & $0.03$ & $0.04$ & $0.02$ & $0.02$ \\
        Charmless background                                       & $0.14$ & $0.15$ & $0.12$ & $0.14$ & $0.01$ & $0.02$ \\
        \CP violation in low-mass background                       & $0.01$ & $0.10$ & $0.08$ & $0.12$ & $0.07$ & $0.26$ \\
        Semi-leptonic $b$-hadron decays                            & $0.05$ & $0.27$ & $0.06$ & $0.01$ & $0.07$ & $0.19$ \\
        Semi-leptonic charm decays                                 & $0.02$ & $0.07$ & $0.03$ & $0.15$ & $0.06$ & $0.24$ \\
        $\D\to \Kmp \pi^\pm \pi^+\pi^-$ background                 & $0.11$ & $0.05$ & $0.07$ & $0.04$ & $0.09$ & $0.05$ \\
        $\Lb \to pD \pi^-$ background                              & $0.01$ & $0.25$ & $0.14$ & $0.04$ & $0.06$ & $0.34$ \\
        $\D\to\kaon^\mp \pion^\pm\pion^+\pion^-\piz$ background    & $0.30$ & $0.05$ & $0.19$ & $0.07$ & $0.05$ & $0.01$ \\
        Fit bias                                                   & $0.06$ & $0.05$ & $0.13$ & $0.02$ & $0.06$ & $0.13$ \\
        \midrule
        Total \lhcb systematic                                     & $0.37$ & $0.43$ & $0.34$ & $0.32$ & $0.70$ & $0.57$ \\
        \midrule
        $c_i$, $s_i$                                               & $0.35$ & $3.64$ & $1.74$ & $1.29$ & $0.14$ & $1.10$ \\
        \midrule
        Total systematic                                           & $0.51$ & $3.67$ & $1.78$ & $1.33$ & $0.72$ & $1.24$ \\
        \midrule
        Statistical                                                & $2.87$ & $3.40$ & $2.51$ & $3.05$ & $4.24$ & $5.17$ \\
        \bottomrule
    \end{tabular}
\end{table}
 
Cross checks are performed with simulation that validate certain assumptions in the analysis, and for which no systematic uncertainties are therefore applied. These studies assess the difference in acceptance over $D$-meson phase space for $B^\pm \to DK^\pm$ and $B^\pm \to D\pi^\pm$ decays, acceptance effects on the effective values of $c_i$ and $s_i$, the effects of bin migration, and the effect of neglecting $\Dz$-$\Dzb$ mixing in the fit. In all cases there are negligible biases on the measured parameters, within the current statistical precision.

The binned measurement is largely insensitive to $\Dz$-$\Dzb$ mixing and bin migration because $F_i$ are free parameters in the fit. Since the mixing and bin-migration effects are very similar between the $\Bpm\to\D\Kpm$ and $\Bpm\to\D\pipm$ modes, the $\Bpm\to\D\pipm$ mode provides a first order correction that is incorporated into the $F_i$ parameters.

 The systematic uncertainties on the phase-space integrated observables are shown in Table~\ref{table:Systematic_uncertainties_GLW}. They are evaluated using the same strategy as those in the binned analysis. In addition, there are systematic uncertainties due to the production and detection asymmetries, which are estimated by repeating the fit many times, each time varying the parameters randomly within their uncertainties~\cite{LHCb-PAPER-2020-036}, and taking the spreads of the resulting distributions as the assigned uncertainties.  The total systematic uncertainties for the asymmetries are an order of magnitude smaller than the statistical uncertainties, and of a similar size for the ratio observables.

\begin{table}[tb]
    \centering
    \caption{Uncertainties on the results of the phase-space integrated analysis.}
    \label{table:Systematic_uncertainties_GLW}
    \begin{tabular}{lcccccc} 
        \toprule
        & \multicolumn{6}{c}{Uncertainty ($\times 10^3$)} \\
        \midrule
        Source & $A_\kaon^{\kaon\kaon\pion\pion}$ & $A_\pion^{\kaon\kaon\pion\pion}$ & $A_\kaon^{\pion\pion\pion\pion}$ & $A_\pion^{\pion\pion\pion\pion}$ & $R_{\CP}^{\kaon\kaon\pion\pion}$ & $R_{\CP}^{\pion\pion\pion\pion}$ \\
        \midrule
        Charmless background                          & $1.2$ & $<0.1$\phantom{00}  & $0.4$ & $<0.1$\phantom{00} & $13.9$ & $8.5$ \\
        External parameters                           & $1.5$ & $0.7$ & $1.5$ & $0.7$ & $4.0$ & $4.0$ \\
        Fixed yield fractions                         & $0.1$ & $<0.1$\phantom{00}  & $0.1$ & $<0.1$\phantom{00}  & $1.3$ & $1.4$ \\
        Mass shape                                    & $0.3$ & $<0.1$\phantom{00}  & $0.2$ & $<0.1$\phantom{00}  & $3.1$ & $3.1$ \\
        PID efficiency                                & $0.1$ & $<0.1$\phantom{00}  & $0.1$ & $<0.1$\phantom{00}  & $2.5$ & $1.6$ \\
        \midrule
        Total systematic                              & $2.0$ & $0.7$ & $1.6$ & $0.7$ & $15.1$ & $10.1$ \\
        \midrule
        Statistical                                   & $23.5$ & $5.5$ & $13.3$ & $3.1$ & $24.2$ & $14.3$ \\
        \bottomrule
    \end{tabular}
\end{table}

Correlation matrices for all the measured observables can be found in Appendix~\ref{section:Correlation_matrices_for_CP_violating_observables}.


\section{Interpretation}
\label{section:Interpretation}
The measured \CP-violating observables in Table~\ref{table:GGSZ_observables} are interpreted in terms of the underlying physics parameters $\gamma$, $\delta_\B^{\D\kaon}$, $r_\B^{\D\kaon}$, $\delta_\B^{\D\pion}$ and $r_\B^{\D\pion}$ using a maximum likelihood fit, following the procedure described in Ref.~\cite{LHCb-PAPER-2021-033}.

\begin{figure}[tb]
    \centering
    \begin{subfigure}{0.5\textwidth}
        \includegraphics[width=1\textwidth]{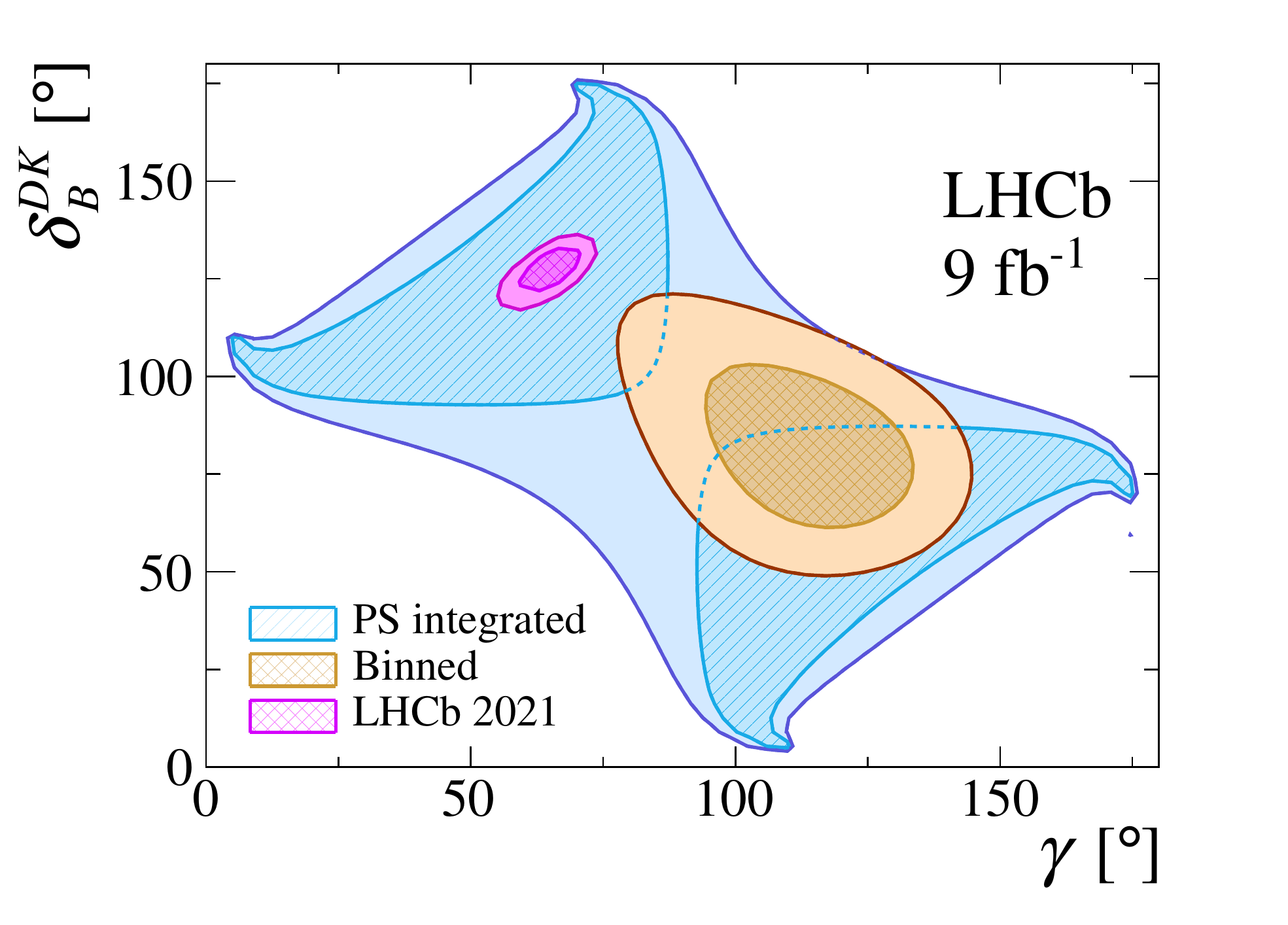}  
    \end{subfigure}%
    \begin{subfigure}{0.5\textwidth}
        \includegraphics[width=1\textwidth]{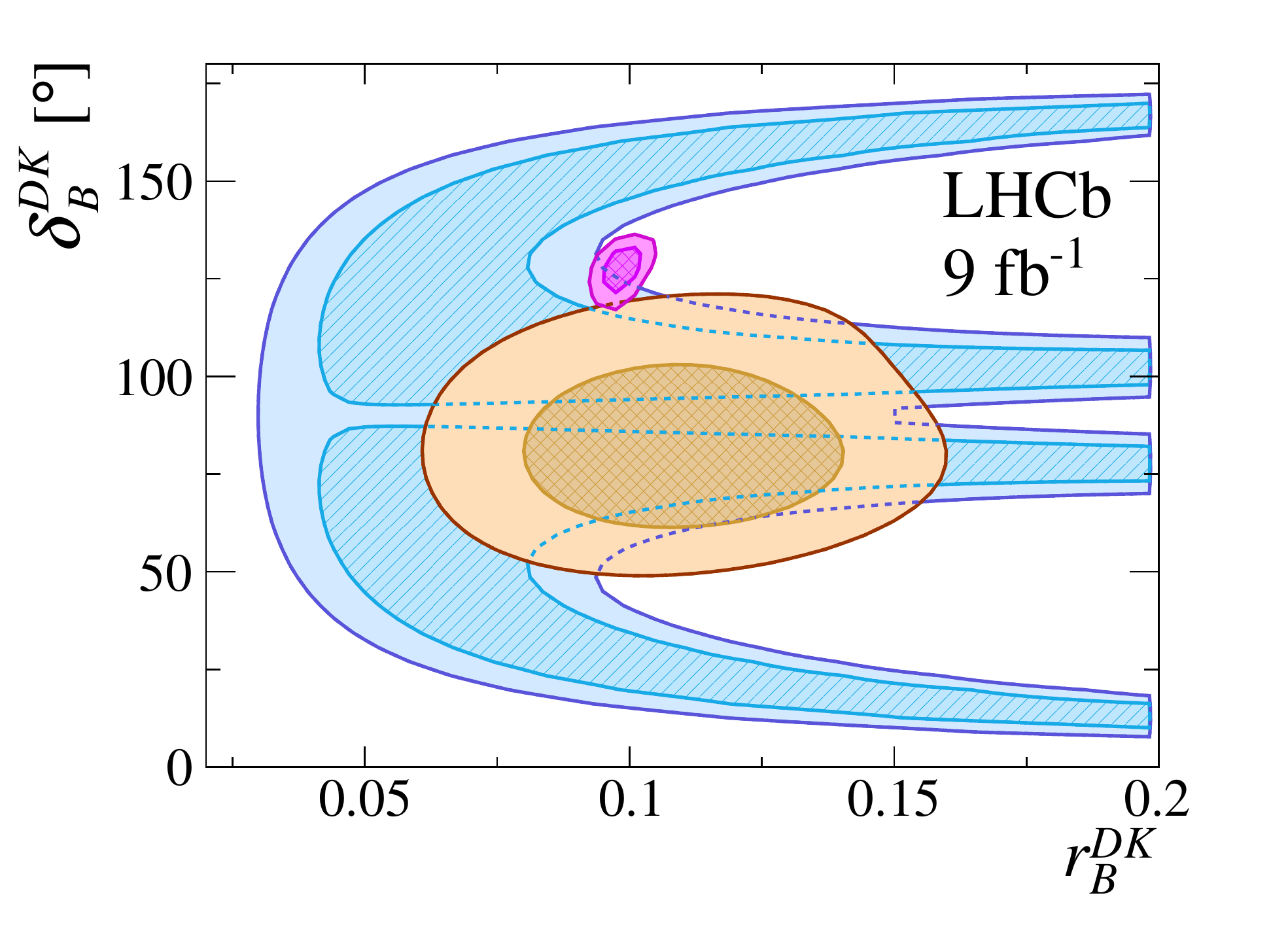}
    \end{subfigure}
    \caption{Interpretation of the binned and phase-space (`PS') integrated measurements in terms to the underlying physics parameters. The 1$\sigma$ and 2$\sigma$ contours are shown, which correspond to $68.3\%$ ($\Delta\chi^2 = 2.30$) and $95.5\%$ ($\Delta\chi^2 = 6.18$) confidence intervals, respectively. Also shown is the result from the analysis of other decay modes at LHCb (`LHCb 2021')~\cite{LHCb-PAPER-2021-033}.}
    \label{figure:GGSZandGLW}
\end{figure}

The fit is first made to the results of the $B^\pm \to [K^+K^-\pi^+\pi^-]_D h^\pm$ binned analysis alone. The 1 and 2$\sigma$ contours in the $\gamma$~vs.~$\delta_B^{DK}$ and the $r_B^{DK}$~vs.~$\delta_B^{DK}$ planes are shown in Fig.~\ref{figure:GGSZandGLW}. The numerical results are
\begin{align*}
    \gamma &= (116^{+12}_{-14})^\circ, \\
    \delta_\B^{\D\kaon} &= (81^{+14}_{-13})^\circ, \\
    r_\B^{\D\kaon} &= 0.110^{+0.020}_{-0.020}, \\
    \delta_\B^{\D\pion} &= (298^{+62}_{-118})^\circ, \\
    r_\B^{\D\pion} &= 0.0041^{+0.0054}_{-0.0041},
\end{align*}
where the uncertainties are the combined statistical and systematic uncertainties, and the results are completely dominated by the former. These model-dependent results may be compared to those from a recent measurement of $\gamma$ and associated parameter derived from an ensemble of beauty and charm-meson decay studies performed by LHCb~\cite{LHCb-PAPER-2021-033}. The $3\sigma$ contours of the results for $\gamma$ and $\delta_B^{DK}$ from the $B^\pm \to [K^+K^-\pi^+\pi^-]_D h^\pm$ analysis encompass the central values coming from the other decay modes.

The fit is then made to the phase-space integrated \CP-violating observables. Here it is necessary to know the \CP-even fractions $F_+$ for each decay, which have been measured by the BESIII collaboration. In the case of $\D \to K^+K^-\pi^+\pi^-$ the measured value is $F_+ = 0.73 \pm 0.04$~\cite{cite:KKpipiFplusBES3} and for $D \to \pi^+\pi^-\pi^+\pi^-$ the value $0.735 \pm 0.016$ is used~\cite{cite:4piFplusBES3}. Due to the trigonometric dependence, multiple solutions are obtained. The likelihood contours are shown in Fig.~\ref{figure:GGSZandGLW}. It can be seen that these are compatible with measurements using other decay channels.


\section{Summary and conclusions}
\label{section:Summary_and_conclusion}
The first measurement of \CP-violating observables for the decay $\Bpm\to[\Kp\Km\pip\pim]_\D h^\pm$ is presented. The analysis is performed in bins of phase space of the $\D$-meson decay, which are chosen to optimise sensitivity to the angle $\gamma$ of the CKM Unitary Triangle. The local asymmetries confirm the presence of \CP violation effects that have also been observed in other $\Bpm\to\D\Kpm$ decay modes. In addition, measurements of \CP-violating observables integrated over phase space are performed for the decays $\Bpm\to[\Kp\Km\pip\pim]_\D h^\pm$ and $\Bpm\to[\pip\pim\pip\pim]_\D h^\pm$. All studies make use of the full data set collected by \lhcb in 2011--2012 and 2015--2018, corresponding to an integrated luminosity of $9\invfb$. The measurement of the phase-space integrated observables in the $\Bpm\to[\pip\pim\pip\pim]_\D h^\pm$ mode supersedes those reported in Ref.~\cite{LHCb-PAPER-2020-036}.

The measurements of the \CP-violating observables in the binned analysis require knowledge of the $\D$-meson strong-phase parameters $c_i$ and $s_i$. The values of these parameters are currently taken from an amplitude model~\cite{LHCb-PAPER-2018-041}. In the future, direct measurements of these parameters at charm threshold~\cite{cite:BESIIIWhitePaper} in combination with the measured yields in each bin, which are reported in Appendix~\ref{section:Yields_in_bins_of_phase_space}, will allow the \CP-violating observables to be determined in a model-independent fashion.

The current measurements, together with those of the integrated charge asymmetries, may be interpreted in terms of $\gamma$ and the other underlying physics parameters. When this is done for the results of the binned $B^\pm \to [K^+K^-\pi^+\pi^-]_D h^\pm$ analysis, a model-dependent value of $\gamma = (116^{+12}_{-14})^\circ$ is obtained. This result will evolve when the observables are re-evaluated using model-independent inputs. A model-independent determination of $\gamma$ making use of both the binned and unbinned analysis of these four-body $D$-meson decay modes will be a valuable addition to the set of measurements of this important parameter already performed at \lhcb. The potential of exploiting modes of this multiplicity for a model-independent measurement of $\gamma$ has already been demonstrated with $\Bpm\to[\Kmp\pipm\pimp\pipm]_\D h^\pm$ decays~\cite{LHCb-PAPER-2022-017}. The precision of the four-body $D$-decay studies is limited by the sample size and is expected to improve significantly with future data from \lhcb.

\section*{Acknowledgements}
%
%
\noindent We express our gratitude to our colleagues in the CERN
accelerator departments for the excellent performance of the LHC. We
thank the technical and administrative staff at the LHCb
institutes.
We acknowledge support from CERN and from the national agencies:
CAPES, CNPq, FAPERJ and FINEP (Brazil); 
MOST and NSFC (China); 
CNRS/IN2P3 (France); 
BMBF, DFG and MPG (Germany); 
INFN (Italy); 
NWO (Netherlands); 
MNiSW and NCN (Poland); 
MEN/IFA (Romania); 
MICINN (Spain); 
SNSF and SER (Switzerland); 
NASU (Ukraine); 
STFC (United Kingdom); 
DOE NP and NSF (USA).
We acknowledge the computing resources that are provided by CERN, IN2P3
(France), KIT and DESY (Germany), INFN (Italy), SURF (Netherlands),
PIC (Spain), GridPP (United Kingdom), 
CSCS (Switzerland), IFIN-HH (Romania), CBPF (Brazil),
Polish WLCG  (Poland) and NERSC (USA).
We are indebted to the communities behind the multiple open-source
software packages on which we depend.
Individual groups or members have received support from
ARC and ARDC (Australia);
Minciencias (Colombia);
AvH Foundation (Germany);
EPLANET, Marie Sk\l{}odowska-Curie Actions and ERC (European Union);
A*MIDEX, ANR, IPhU and Labex P2IO, and R\'{e}gion Auvergne-Rh\^{o}ne-Alpes (France);
Key Research Program of Frontier Sciences of CAS, CAS PIFI, CAS CCEPP, 
Fundamental Research Funds for the Central Universities, 
and Sci. \& Tech. Program of Guangzhou (China);
GVA, XuntaGal, GENCAT and Prog. Atracci\'on Talento, CM (Spain);
SRC (Sweden);
the Leverhulme Trust, the Royal Society
 and UKRI (United Kingdom).



\section*{Appendices}

\appendix

\section{Yields in bins of phase space and alternative binning scheme}
\label{section:Yields_in_bins_of_phase_space}

The fitted parameters $x_\pm^{\D\kaon}$, $y_\pm^{\D\kaon}$, $x_\xi^{\D\pion}$ and $y_\xi^{\D\pion}$ are determined using values of $c_i$ and $s_i$ calculated from an amplitude model. When direct measurements of $c_i$ and $s_i$ are available from studies at charm threshold it will be desirable to update the $\Bpm\to[\Kp\Km\pip\pim]_\D h^\pm$ binned analysis with this information as input. In order to enable this re-analysis an alternative fit is carried out where the yields in each bin are fitted, instead of the \CP-violating observables. The bin yields for the $2 \times 8$ scheme are presented in Sec.~\ref{subsection:2x8_bins}.

It is possible that the sample sizes at charm threshold may be smaller than foreseen.  In that case a $2 \times 4$ binning scheme might be more suited to the analysis. Therefore, in this Appendix a $2 \times 4$ binning scheme is presented in Sec.~\ref{subsection:2x4_bins} together with the bin yields and essential information for this scheme. The corresponding correlation matrices are available in Ref.~\cite{cite:KKpipiBinYields}.

\subsection{\texorpdfstring{\boldmath{$2\times 8$}}{2x8} bins}
\label{subsection:2x8_bins}

Table~\ref{table:Bin_yields_2x8} lists the bin yields for the $2\times 8$ binning scheme. When the \CP observables are determined from the bin yields, rather than as an output of the fit to the invariant-mass spectrum, the fit biases change.  The biases with this new configuration, which should be used to correct the observables and replace the corresponding systematic uncertainty in  Table~\ref{table:Systematic_uncertainties}, are presented in Table~\ref{table:2Step_fit_bias}. The new correlation matrix for systematic uncertainties is presented in Table~\ref{table:GGSZ_fit_correlations_systematic_2step}, which only contains contributions from internal systematic uncertainties.

\begin{table}[tb]
    \centering
    \caption{Yields with the $2\times 8$ binning scheme.}
    \label{table:Bin_yields_2x8}
    \begin{tabular}{ccccc} 
        \toprule
        Bin             & $\Bm\to\D\Km$   & $\Bp\to\D\Kp$   & $\Bm\to\D\pim$  & $\Bp\to\D\pip$  \\
        \midrule
        $\phantom{+}8$             & $\phantom{0}17 \pm 6\phantom{0}$      & $\phantom{0}74 \pm 10$                & $\phantom{0}312 \pm 21$    & $\phantom{0}920 \pm 34$    \\
        $\phantom{+}7$             & $\phantom{0}21 \pm 7\phantom{0}$      & $\phantom{0}71 \pm 10$                & $\phantom{0}309 \pm 21$    & $1160 \pm 37$   \\
        $\phantom{+}6$             & $\phantom{0}81 \pm 12$                & $173 \pm 15$                          & $1025 \pm 36$              & $2422 \pm 53$   \\
        $\phantom{+}5$             & $157 \pm 15$                          & $271 \pm 19$                          & $2103 \pm 50$              & $4226 \pm 68$   \\
        $\phantom{+}4$             & $146 \pm 15$                          & $230 \pm 17$                          & $1750 \pm 46$              & $3899 \pm 66$   \\
        $\phantom{+}3$             & $\phantom{0}52 \pm 9\phantom{0}$      & $143 \pm 14$                          & $\phantom{0}671 \pm 30$    & $2554 \pm 54$   \\
        $\phantom{+}2$             & $\phantom{0}43 \pm 9\phantom{0}$      & $120 \pm 13$                          & $\phantom{0}468 \pm 25$    & $1417 \pm 41$   \\
        $\phantom{+}1$             & $\phantom{0}11 \pm 6\phantom{0}$      & $\phantom{0}65 \pm 10$                & $\phantom{0}369 \pm 22$    & $1137 \pm 37$   \\
        $-1$                       & $\phantom{0}66 \pm 10$                & $\phantom{0}26 \pm 7\phantom{0}$      & $1009 \pm 35$              & $\phantom{0}376 \pm 23$    \\
        $-2$                       & $\phantom{0}93 \pm 12$                & $\phantom{0}51 \pm 9\phantom{0}$      & $1477 \pm 41$              & $\phantom{0}442 \pm 25$    \\
        $-3$                       & $152 \pm 15$                          & $\phantom{0}39 \pm 9\phantom{0}$      & $2424 \pm 53$              & $\phantom{0}690 \pm 30$    \\
        $-4$                       & $277 \pm 19$                          & $\phantom{0}88 \pm 12$                & $3800 \pm 65$              & $1851 \pm 47$   \\
        $-5$                       & $339 \pm 21$                          & $\phantom{0}93 \pm 13$                & $4185 \pm 68$              & $2210 \pm 50$   \\
        $-6$                       & $180 \pm 15$                          & $\phantom{0}46 \pm 9\phantom{0}$      & $2375 \pm 52$              & $\phantom{0}939 \pm 34$    \\
        $-7$                       & $\phantom{0}61 \pm 10$                & $\phantom{0}34 \pm 8\phantom{0}$      & $1127 \pm 36$              & $\phantom{0}376 \pm 23$    \\
        $-8$                       & $\phantom{0}71 \pm 10$                & $\phantom{0}29 \pm 7\phantom{0}$      & $\phantom{0}987 \pm 34$    & $\phantom{0}283 \pm 20$    \\
        \bottomrule
    \end{tabular}
\end{table}

\begin{table}[tb]
    \centering
    \caption{Fit bias when using the bin yields for the $2\times 8$ binning scheme.}
    \label{table:2Step_fit_bias}
    \begin{tabular}{cccccc}
        \toprule
        \multicolumn{6}{c}{Bias ($\times 10^2$)} \\
        \midrule
        $x_-^{\D\kaon}$ & $y_-^{\D\kaon}$ & $x_+^{\D\kaon}$ & $y_+^{\D\kaon}$ & $x_\xi^{\D\pion}$ & $y_\xi^{\D\pion}$ \\
        \midrule
        $0.26$ & $0.11$ & $0.12$ & $0.04$ & $0.21$ & $0.10$ \\
        \bottomrule
    \end{tabular}
\end{table}

\begin{table}[tb]
    \centering
    \caption{Correlation matrix for systematic uncertainties of \CP-violating observables for the binned measurement, with the $2\times 8$ binning scheme. The contribution from $c_i$ and $s_i$ is excluded.}
    \label{table:GGSZ_fit_correlations_systematic_2step}
    \begin{tabular}{crrrrrr} 
        \toprule
        & $x_-^{\D\kaon}$ & $y_-^{\D\kaon}$ & $x_+^{\D\kaon}$ & $y_+^{\D\kaon}$ & $x_\xi^{\D\pion}$ & $y_\xi^{\D\pion}$ \\
        \midrule
        $x_-^{\D\kaon}$ & $1.000$  & $0.010$  & $0.572$  & $0.411$  & $-0.258$ & $-0.127$ \\
        $y_-^{\D\kaon}$ & $0.010$  & $1.000$  & $0.099$  & $0.083$  & $-0.010$ & $0.282$  \\
        $x_+^{\D\kaon}$ & $0.572$  & $0.099$  & $1.000$  & $0.397$  & $-0.158$ & $0.119$  \\
        $y_+^{\D\kaon}$ & $0.411$  & $0.083$  & $0.397$  & $1.000$  & $-0.541$ & $-0.461$ \\
        $x_\xi^{\D\pion}$ & $-0.258$ & $-0.010$ & $-0.158$ & $-0.541$ & $1.000$  & $0.269$  \\
        $y_\xi^{\D\pion}$ & $-0.127$ & $0.282$  & $0.119$  & $-0.461$ & $0.269$  & $1.000$  \\
        \bottomrule
    \end{tabular}
\end{table}

\subsection{\texorpdfstring{\boldmath{$2\times 4$}}{2x4} bins}
\label{subsection:2x4_bins}

\begin{figure}[tb]
    \centering
    \begin{subfigure}{0.57\textwidth}
        \includegraphics[height=6.2cm]{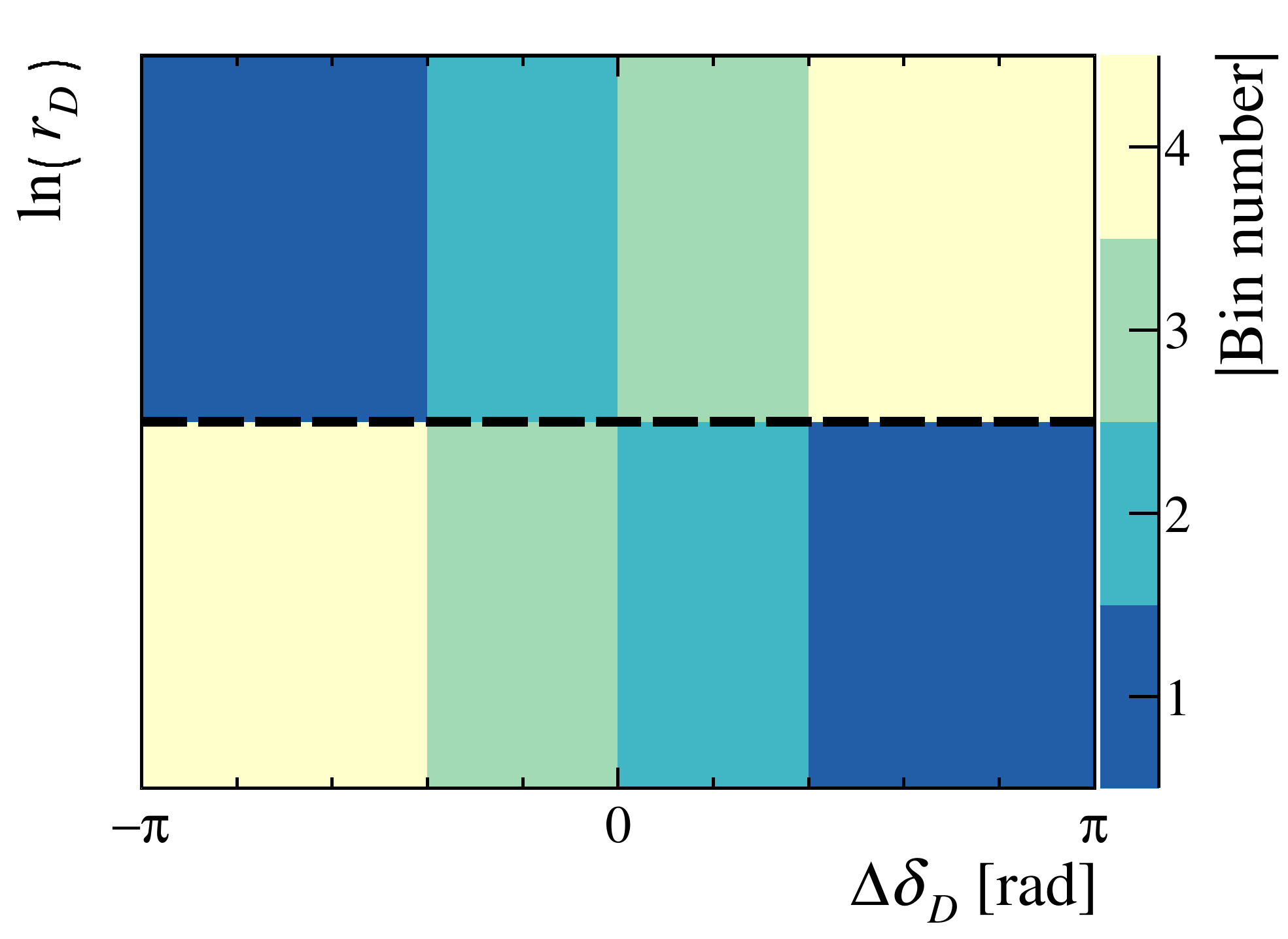}
    \end{subfigure}%
    \hfill
    \begin{subfigure}{0.43\textwidth}
        \includegraphics[height=6.2cm]{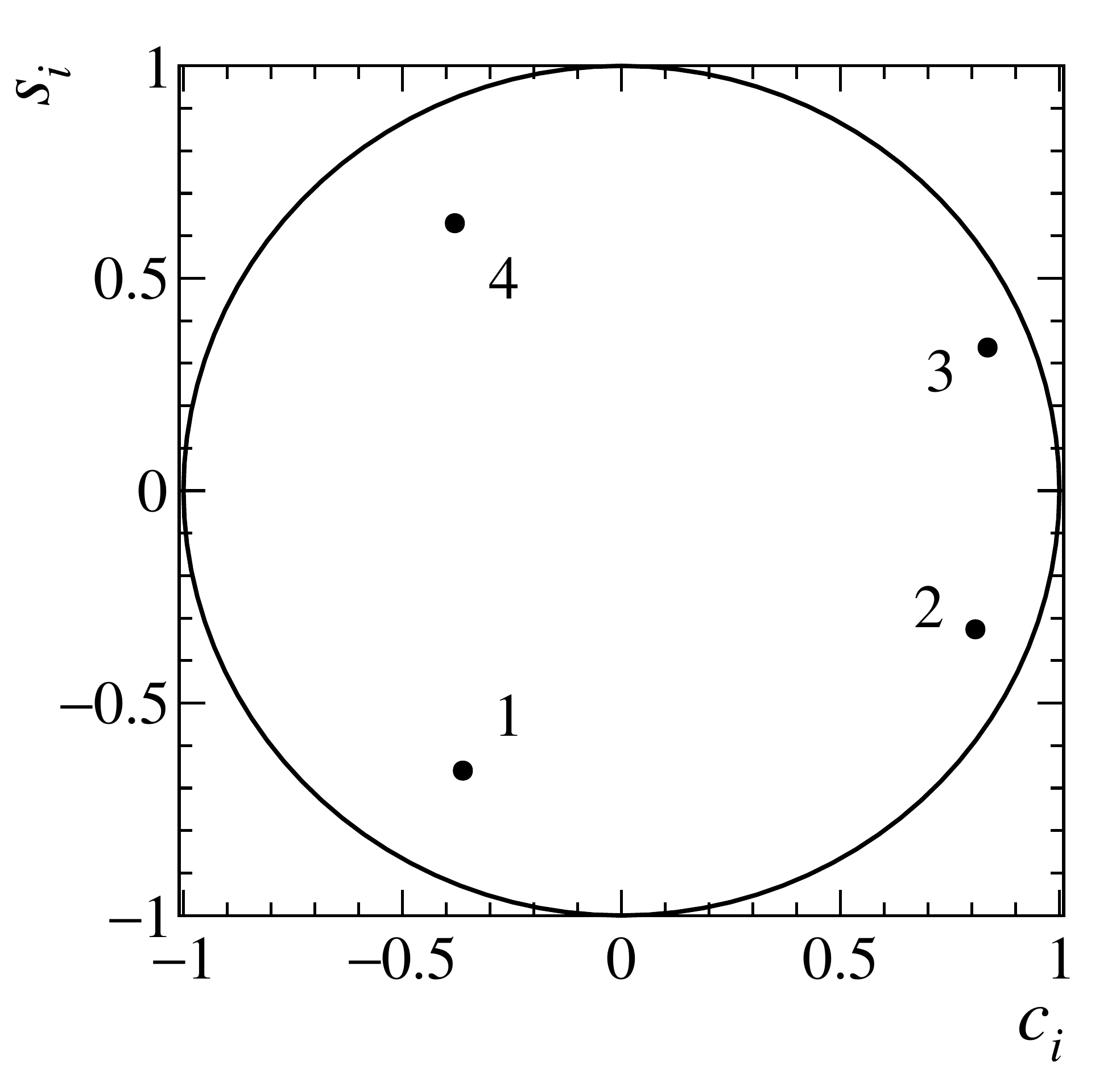}
    \end{subfigure}
    \caption{Left: Optimised $2\times 4$ binning scheme in $\Delta \delta_D$-$\ln(r_D)$ space. Right: The associated $c_i$ and $s_i$ parameters calculated using the amplitude model (right). The numbers indicate the bin numbers.}
    \label{figure:Binning_scheme_plots_4bins}
\end{figure}

The $2\times 4$ binning scheme is presented in Fig.~\ref{figure:Binning_scheme_plots_4bins}, and the corresponding $c_i$, $s_i$ and $F_i$ parameters, and the normalised bin volumes $V_i$ calculated from the model, are listed in Table~\ref{table:ci_si_Fi_2x4}. The optimised $Q$-value is $Q = 0.85$. The code provided in Ref.~\cite{cite:KKpipiBinningScheme} can be used to assign bin numbers to $\D$ decays with this binning scheme as well.

\begin{table}[tb]
    \centering
    \caption{Values of $c_i$, $s_i$, $F_i$ and $V_i$ for the optimised $2\times 4$ binning scheme, as calculated from the amplitude model.}
    \label{table:ci_si_Fi_2x4}
    \begin{tabular}{crrccccc}
        \toprule
        Bin number& $c_i$     & $s_i$     & $F_i$     & $F_{-i}$& $V_i$     & $V_{-i}$ \\
        \midrule
        $1$       & $-0.3623$ & $-0.6585$ & $0.0327$  & $0.1085$  & $0.1152$  & $0.1153$   \\
        $2$       & $0.8038$  & $-0.3325$ & $0.0997$  & $0.2544$  & $0.1455$  & $0.1455$   \\
        $3$       & $0.8341$  & $0.3387$  & $0.1225$  & $0.2656$  & $0.1442$  & $0.1442$   \\
        $4$       & $-0.3862$ & $0.6269$  & $0.0263$  & $0.0902$  & $0.0950$  & $0.0950$   \\
        \bottomrule
    \end{tabular}
\end{table}

Table~\ref{table:Bin_yields_2x4} lists the bin yields for the $2\times 4$ binning scheme. Additionally, the internal systematic uncertainties, evaluated for a $2\times 4$ binning scheme, are also provided in Table~\ref{table:Systematic_uncertainties_4Bins}, which are evaluated in an identical manner to those found in Table~\ref{table:Systematic_uncertainties}.  The correlation matrix of the systematic uncertainties is given in Table~\ref{table:GGSZ_fit_correlations_systematic_4Bins}.

\begin{table}[tb]
    \centering
    \caption{Yields with the $2\times 4$ binning scheme.}
    \label{table:Bin_yields_2x4}
    \begin{tabular}{ccccc} 
        \toprule
        Bin             & $\Bm\to\D\Km$   & $\Bp\to\D\Kp$   & $\Bm\to\D\pim$  & $\Bp\to\D\pip$  \\
        \midrule
        $\phantom{+}4$             & $\phantom{0}36 \pm 9\phantom{0}$      & $140 \pm 14$               & $\phantom{0}600 \pm 29$    & $1969 \pm 49$   \\
        $\phantom{+}3$             & $240 \pm 19$                          & $448 \pm 25$               & $3140 \pm 62$              & $6753 \pm 87$   \\
        $\phantom{+}2$             & $197 \pm 18$                          & $374 \pm 23$               & $2452 \pm 55$              & $6561 \pm 86$   \\
        $\phantom{+}1$             & $\phantom{0}50 \pm 10$                & $181 \pm 16$               & $\phantom{0}799 \pm 33$    & $2436 \pm 54$   \\
        $-1$                       & $153 \pm 15$                          & $\phantom{0}76 \pm 12$     & $2376 \pm 53$              & $\phantom{0}792 \pm 33$    \\
        $-2$                       & $436 \pm 25$                          & $130 \pm 15$               & $6332 \pm 85$              & $2559 \pm 56$   \\
        $-3$                       & $528 \pm 26$                          & $142 \pm 16$               & $6649 \pm 87$              & $3177 \pm 62$   \\
        $-4$                       & $122 \pm 14$                          & $\phantom{0}60 \pm 10$     & $2022 \pm 49$              & $\phantom{0}626 \pm 30$    \\
        \bottomrule
    \end{tabular}
\end{table}

\begin{table}[tb]
    \centering
    \caption{Systematic uncertainties on the results of the binned analysis, with the $2\times 4$ binning scheme.}
    \label{table:Systematic_uncertainties_4Bins}
    \begin{tabular}{lcccccc}
        \toprule
        & \multicolumn{6}{c}{Uncertainty ($\times 10^2$)} \\
        \midrule
        Source & $x_-^{\D\kaon}$ & $y_-^{\D\kaon}$ & $x_+^{\D\kaon}$ & $y_+^{\D\kaon}$ & $x_\xi^{\D\pion}$ & $y_\xi^{\D\pion}$ \\
        \midrule
        Mass shape                                              & $0.01$ & $0.01$ & $0.02$ & $0.04$ & $0.02$ & $0.03$ \\
        Bin-dependent mass shape                                & $0.31$ & $0.43$ & $0.36$ & $0.10$ & $0.40$ & $0.01$ \\
        PID efficiency                                          & $0.01$ & $0.01$ & $0.02$ & $0.04$ & $0.02$ & $0.03$ \\
        Low-mass background model                               & $0.01$ & $0.01$ & $0.01$ & $0.01$ & $0.03$ & $0.01$ \\
        Charmless background                                    & $0.11$ & $0.17$ & $0.09$ & $0.14$ & $0.01$ & $0.01$ \\
        \CP violation in low-mass background                    & $0.05$ & $0.22$ & $0.07$ & $0.14$ & $0.15$ & $0.13$ \\
        Semi-leptonic $b$-hadron decays                         & $0.07$ & $0.08$ & $0.03$ & $0.07$ & $0.17$ & $0.56$ \\
        Semi-leptonic charm decays                              & $0.02$ & $0.01$ & $0.04$ & $0.02$ & $0.15$ & $0.29$ \\
        $\D\to\kaon^\mp\pion^\pm\pion^+\pion^-$ background      & $0.02$ & $0.09$ & $0.01$ & $0.06$ & $0.25$ & $0.61$ \\
        $\Lb \to p D \pi^-$ background                          & $0.16$ & $0.04$ & $0.13$ & $0.04$ & $0.01$ & $0.12$ \\
        $\D\to\kaon^\mp\pion^\pm\pion^+\pion^-\piz$ background  & $0.09$ & $0.11$ & $0.44$ & $0.08$ & $0.12$ & $0.46$ \\
        Fit bias                                                & $0.19$ & $0.27$ & $0.07$ & $0.19$ & $0.51$ & $0.21$ \\
        \midrule
        Total \lhcb systematic                                  & $0.43$ & $0.60$ & $0.61$ & $0.32$ & $0.76$ & $1.03$ \\
        \bottomrule
    \end{tabular}
\end{table}

\begin{table}[tb]
    \centering
    \caption{Correlation matrix for systematic uncertainties of \CP-violating observables for the binned measurement, with the $2\times 4$ binning scheme. The contribution from $c_i$ and $s_i$ is excluded.}
    \label{table:GGSZ_fit_correlations_systematic_4Bins}
    \begin{tabular}{crrrrrr} 
        \toprule
        & $x_-^{\D\kaon}$ & $y_-^{\D\kaon}$ & $x_+^{\D\kaon}$ & $y_+^{\D\kaon}$ & $x_\xi^{\D\pion}$ & $y_\xi^{\D\pion}$ \\
        \midrule
        $x_-^{\D\kaon}$ & $1.000$  & $-0.591$ & $0.742$  & $-0.329$ & $-0.504$ & $0.290$  \\
        $y_-^{\D\kaon}$ & $-0.591$ & $1.000$  & $-0.388$ & $0.622$  & $0.388$  & $0.036$  \\
        $x_+^{\D\kaon}$ & $0.742$  & $-0.388$ & $1.000$  & $-0.586$ & $-0.250$ & $0.355$  \\
        $y_+^{\D\kaon}$ & $-0.329$ & $0.622$  & $-0.586$ & $1.000$  & $0.067$  & $0.000$  \\
        $x_\xi^{\D\pion}$ & $-0.504$ & $0.388$  & $-0.250$ & $0.067$  & $1.000$  & $-0.043$ \\
        $y_\xi^{\D\pion}$ & $0.290$  & $0.036$  & $0.355$  & $0.000$  & $-0.043$ & $1.000$  \\
        \bottomrule
    \end{tabular}
\end{table}

\section{Correlation matrices for \texorpdfstring{\boldmath{\CP}}{CP}-violating observables}
\label{section:Correlation_matrices_for_CP_violating_observables}
The statistical and systematic correlation matrices of the \CP-violating observables from the binned and phase-space integrated measurements are shown in Tables~\ref{table:2step_CP_fit_correlations}-\ref{table:GLW_fit_correlations_systematic}. Tables~\ref{table:2step_CP_fit_correlations} and \ref{table:GGSZ_fit_correlations_systematic} include contributions from $c_i$ and $s_i$, which are currently model dependent.

\begin{table}[tb]
    \centering
    \caption{Correlation matrix for statistical uncertainties of \CP-violating observables for the binned measurement.}
    \label{table:2step_CP_fit_correlations}
    \begin{tabular}{crrrrrr} 
        \toprule
        & $x_-^{\D\kaon}$ & $y_-^{\D\kaon}$ & $x_+^{\D\kaon}$ & $y_+^{\D\kaon}$ & $x_\xi^{\D\pion}$ & $y_\xi^{\D\pion}$ \\
        \midrule
        $x_-^{\D\kaon}$ & $1.000$  & $0.032$  & $0.008$  & $-0.010$ & $0.034$  & $0.102$  \\
        $y_-^{\D\kaon}$ & $0.032$  & $1.000$  & $0.017$  & $0.000$  & $-0.091$ & $0.080$  \\
        $x_+^{\D\kaon}$ & $0.008$  & $0.017$  & $1.000$  & $0.007$  & $-0.100$ & $0.051$  \\
        $y_+^{\D\kaon}$ & $-0.010$ & $0.000$  & $0.007$  & $1.000$  & $0.012$  & $-0.097$ \\
        $x_\xi^{\D\pion}$ & $0.034$  & $-0.091$ & $-0.100$ & $0.012$  & $1.000$  & $0.014$  \\
        $y_\xi^{\D\pion}$ & $0.102$  & $0.080$  & $0.051$  & $-0.097$ & $0.014$  & $1.000$  \\
        \bottomrule
    \end{tabular}
\end{table}

\begin{table}[tb]
    \centering
    \caption{Correlation matrix for systematic uncertainties of \CP-violating observables for the binned measurement.}
    \label{table:GGSZ_fit_correlations_systematic}
    \begin{tabular}{crrrrrr} 
        \toprule
        & $x_-^{\D\kaon}$ & $y_-^{\D\kaon}$ & $x_+^{\D\kaon}$ & $y_+^{\D\kaon}$ & $x_\xi^{\D\pion}$ & $y_\xi^{\D\pion}$ \\
        \midrule
        $x_-^{\D\kaon}$ & $1.000$  & $-0.678$ & $0.751$  & $0.736$  & $-0.048$ & $-0.650$ \\
        $y_-^{\D\kaon}$ & $-0.678$ & $1.000$  & $-0.973$ & $-0.961$ & $-0.200$ & $0.898$  \\
        $x_+^{\D\kaon}$ & $0.751$  & $-0.973$ & $1.000$  & $0.971$  & $0.166$  & $-0.862$ \\
        $y_+^{\D\kaon}$ & $0.736$  & $-0.961$ & $0.971$  & $1.000$  & $0.065$  & $-0.913$ \\
        $x_\xi^{\D\pion}$ & $-0.048$ & $-0.200$ & $0.166$  & $0.065$  & $1.000$  & $-0.057$ \\
        $y_\xi^{\D\pion}$ & $-0.650$ & $0.898$  & $-0.862$ & $-0.913$ & $-0.057$ & $1.000$  \\
        \bottomrule
    \end{tabular}
\end{table}

\begin{table}[tb]
    \centering
    \caption{Correlation matrix for statistical uncertainties of \CP-violating observables for the phase-space integrated measurement.}
    \label{table:GLW_fit_correlations}
    \begin{tabular}{crrrrrr} 
        \toprule
        & $A_{\kaon}^{\kaon\kaon\pion\pion}$ & $A_{\pion}^{\kaon\kaon\pion\pion}$ & $A_{\kaon}^{\pion\pion\pion\pion}$ & $A_{\pion}^{\pion\pion\pion\pion}$ & $R_{\rm CP}^{\kaon\kaon\pion\pion}$ & $R_{\rm CP}^{\pion\pion\pion\pion}$ \\
        \midrule
        $A_{\kaon}^{\kaon\kaon\pion\pion}$ & $1.000$  & $-0.025$ & $0.000$  & $0.000$  & $0.015$  & $0.000$  \\
        $A_{\pion}^{\kaon\kaon\pion\pion}$ & $-0.025$ & $1.000$  & $0.000$  & $0.000$  & $0.002$  & $0.000$  \\
        $A_{\kaon}^{\pion\pion\pion\pion}$ & $0.000$  & $0.000$  & $1.000$  & $0.028$  & $0.002$  & $0.016$  \\
        $A_{\pion}^{\pion\pion\pion\pion}$ & $0.000$  & $0.000$  & $-0.028$ & $1.000$  & $0.000$  & $0.002$  \\
        $R_{\rm CP}^{\kaon\kaon\pion\pion}$ & $0.015$  & $0.002$  & $0.002$  & $0.000$  & $1.000$  & $0.068$  \\
        $R_{\rm CP}^{\pion\pion\pion\pion}$ & $0.000$  & $0.000$  & $0.016$  & $0.002$  & $0.068$  & $1.000$  \\
        \bottomrule
    \end{tabular}
\end{table}

\begin{table}[tb]
    \centering
    \caption{Correlation matrix for systematic uncertainties of \CP-violating observables for the phase-space integrated measurement.}
    \label{table:GLW_fit_correlations_systematic}
    \begin{tabular}{crrrrrr}
        \toprule
        & $A_\kaon^{\kaon\kaon\pion\pion}$ & $A_\pion^{\kaon\kaon\pion\pion}$ & $A_\kaon^{\pion\pion\pion\pion}$ & $A_\pion^{\pion\pion\pion\pion}$ & $R_{\CP}^{\kaon\kaon\pion\pion}$ & $R_{\CP}^{\pion\pion\pion\pion}$ \\
        \midrule
        $A_\kaon^{\kaon\kaon\pion\pion}$ & $1.000$  & $0.345$  & $0.758$  & $0.343$  & $-0.589$ & $-0.056$ \\
        $A_\pion^{\kaon\kaon\pion\pion}$ & $0.345$  & $1.000$  & $0.423$  & $0.999$  & $0.002$  & $0.007$  \\
        $A_\kaon^{\pion\pion\pion\pion}$ & $0.758$  & $0.423$  & $1.000$  & $0.420$  & $-0.033$ & $-0.246$ \\
        $A_\pion^{\pion\pion\pion\pion}$ & $0.343$  & $0.999$  & $0.420$  & $1.000$  & $0.009$  & $0.014$  \\
        $R_{\CP}^{\kaon\kaon\pion\pion}$ & $-0.589$ & $0.002$  & $-0.033$ & $0.009$  & $1.000$  & $0.236$  \\
        $R_{\CP}^{\pion\pion\pion\pion}$ & $-0.056$ & $0.007$  & $-0.246$ & $0.014$  & $0.236$  & $1.000$  \\
        \bottomrule
    \end{tabular}
\end{table}

\clearpage


\addcontentsline{toc}{section}{References}
\bibliographystyle{LHCb}
\bibliography{main,standard,LHCb-PAPER,LHCb-CONF,LHCb-DP,LHCb-TDR}

\newpage
\centerline
{\large\bf LHCb collaboration}
\begin
{flushleft}
\small
R.~Aaij$^{32}$\lhcborcid{0000-0003-0533-1952},
A.S.W.~Abdelmotteleb$^{50}$\lhcborcid{0000-0001-7905-0542},
C.~Abellan~Beteta$^{44}$,
F.~Abudin{\'e}n$^{50}$\lhcborcid{0000-0002-6737-3528},
T.~Ackernley$^{54}$\lhcborcid{0000-0002-5951-3498},
B.~Adeva$^{40}$\lhcborcid{0000-0001-9756-3712},
M.~Adinolfi$^{48}$\lhcborcid{0000-0002-1326-1264},
P.~Adlarson$^{77}$\lhcborcid{0000-0001-6280-3851},
H.~Afsharnia$^{9}$,
C.~Agapopoulou$^{13}$\lhcborcid{0000-0002-2368-0147},
C.A.~Aidala$^{78}$\lhcborcid{0000-0001-9540-4988}
% ...

{\footnotesize \it

% ...
$^{9}$Universit{\'e} Clermont Auvergne, CNRS/IN2P3, LPC, Clermont-Ferrand, France\\
% ...
$^{13}$LPNHE, Sorbonne Universit{\'e}, Paris Diderot Sorbonne Paris Cit{\'e}, CNRS/IN2P3, Paris, France\\
% ...
$^{32}$Nikhef National Institute for Subatomic Physics, Amsterdam, Netherlands\\
% ...
$^{40}$Instituto Galego de F{\'\i}sica de Altas Enerx{\'\i}as (IGFAE), Universidade de Santiago de Compostela, Santiago de Compostela, Spain\\
% ...
$^{44}$Physik-Institut, Universit{\"a}t Z{\"u}rich, Z{\"u}rich, Switzerland\\
% ...
$^{48}$H.H. Wills Physics Laboratory, University of Bristol, Bristol, United Kingdom\\
% ...
$^{50}$Department of Physics, University of Warwick, Coventry, United Kingdom\\
% ...
$^{53}$School of Physics and Astronomy, University of Glasgow, Glasgow, United Kingdom\\
$^{54}$Oliver Lodge Laboratory, University of Liverpool, Liverpool, United Kingdom\\
% ...
$^{62}$Syracuse University, Syracuse, NY, United States\\
% ...
$^{77}$Department of Physics and Astronomy, Uppsala University, Uppsala, Sweden, associated to $^{53}$\\
$^{78}$University of Michigan, Ann Arbor, MI, United States, associated to $^{62}$\\
\bigskip
% ...
}
\end{flushleft}

\end{document}